\documentstyle[12pt]{article}

\makeatletter
\def\section{\@startsection {section}{1}{\z@}{-3.5ex plus -1ex minus
 -.2ex}{2.3ex plus .2ex}{\large\bf}}
\def\subsection{\@startsection{subsection}{2}{\z@}{-3.25ex plus -1ex minus
 -.2ex}{1.5ex plus .2ex}{\normalsize\bf}}
\makeatother

\begin{document}
\global\parskip=4pt

%%%%%%%%%%%%%%%%%%%%%%%%%%%%%%%%%%%%%%%%%%%%%%%%%%%%%%%%%%%%%%%%%%%%%%%%%%%%

%%HERE ARE SOME GENERAL DEFINITIONS AND COMMANDS
\newcommand{\nc}{\newcommand}
\newcommand{\rnc}{\renewcommand}

%%this numbers equations by sections
\makeatletter
\@addtoreset{equation}{section}
\rnc{\theequation}{\arabic{section}.\arabic{equation}}
\makeatother

%%this ignores (comments out) the text in {}s
%%(in a not particularly elegant way)
%%usage: \ignore{text to be ignored}
\nc{\ignore}[1]{}

\nc{\be}{\begin{equation}}
\nc{\ee}{\end{equation}}
\nc{\bea}{\begin{eqnarray}}
\nc{\eea}{\end{eqnarray}}

%%GREEK LETTERS
\rnc{\a}{\alpha}
\rnc{\d}{\delta}
\nc{\e}{\eta}
\nc{\eb}{\bar{\eta}}
\nc{\f}{\phi}
\nc{\fb}{\bar{\phi}}
\nc{\vf}{\varphi}
\nc{\p}{\psi}
\rnc{\pb}{\bar{\psi}}
\rnc{\c}{\chi}
\nc{\cb}{\bar{\c}}
\nc{\la}{\lambda}
\nc{\m}{\mu}
\nc{\n}{\nu}
\rnc{\o}{\omega}
\rnc{\t}{\theta}
\nc{\tb}{\bar{\theta}}
\nc{\eps}{\epsilon}

%%VARIOUS MANIFOLDS
\rnc{\S}{\Sigma}
\nc{\Sa}{\S\times\{0\}}
\nc{\Sb}{\S\times\{1\}}
\nc{\SI}{\S\times I}
\nc{\SS}{\S\times S^{1}}
\nc{\M}{{\cal M}}

%%OTHER USEFUL ABBREVIATIONS

%%this produces textstyle fractions in displayed equations
\nc{\trac}[2]{{\textstyle\frac{#1}{#2}}}

%%2x2 matrices
\nc{\mat}[4]{\left(\begin{array}{cc}#1&#2\\#3&#4\end{array}\right)}

\def\tr{\mathop{\rm tr}\nolimits}
\def\Tr{\mathop{\rm Tr}\nolimits}
\nc{\ra}{\rightarrow}
\nc{\ot}{\otimes}
\rnc{\ss}{\subset}

%%GAUGE FIELDS, INTEGRALS, AND SUCH

\rnc{\lg}{{\bf g}}
\nc{\lt}{{\bf t}}
\nc{\lk}{{\bf k}}
\nc{\del}{\partial}
\nc{\dz}{\del_{z}}
\nc{\dzb}{\del_{\bar{z}}}
\nc{\az}{A_{z}}
\nc{\azb}{A_{\bar{z}}}
\nc{\bz}{B_{z}}
\nc{\bzb}{B_{\bar{z}}}
\nc{\ba}{{\bf A}}
\nc{\bb}{{\bf B}}
\nc{\g}{g^{-1}}

%\nc{\idl}{\int_{0}^{1}\!d\la\;}
%\nc{\idt}[2]{\int_{#1}^{#2}\!dt\;}
%\nc{\elf}{e^{i\la\f}}
%\nc{\emlf}{e^{-i\la\f}}
%\nc{\etf}{e^{it\f}}
%\nc{\emtf}{e^{-it\f}}

\nc{\dw}{\Delta_{W}}% the weyl determinant
\nc{\Det}{{\rm Det}\,} %functional \det
\nc{\bG}{{\bf G}}
\nc{\bT}{{\bf T}}
\nc{\unA}{\underline{A}}                    %UNDERLINES
\nc{\bft}{{\bf t}}                          %BOLD FACES
\nc{\bfk}{{\bf k}}
\nc{\bfg}{{\bf g}}
\nc{\C}{{\cal A}/{\cal G}}                  %SPACE OF GAUGE ORBITS
\nc{\A}[1]{{\cal A}^{#1}/{\cal G}^{#1}}     %SPACE OF GAUGE ORBITS
\nc{\dx}{\dot{x}}
\rnc{\O}[2]{\Omega^{#1}({#2},\lg)}          %LIE ALGEBRA VALUED FORMS
%%%%%%%%%%%%%%%%%%%%%%%%%%%%%%%%%%%%%%%%%%%%%%%%%%%%%%%%%%%%%

%%LASTLY SOME MORE TIME-SAVERS
\nc{\CS}{Chern-Simons}
\def\tft{top\-o\-log\-i\-cal field the\-o\-ry}
\def\tgt{top\-o\-log\-i\-cal gauge the\-o\-ry}
\def\wif{Weyl int\-egral form\-ula}
\def\ver{Ver\-lin\-de form\-ula}
%%%%%%%%%%%%%%%%%%%%%%%%%%%%%%%%%%%%%%%%%%%%%%%%%%%%%%%%%%%%%

\begin{titlepage}
\newlength{\titlehead}
\settowidth{\titlehead}{NIKHEF-H/91}
\begin{flushright}
\parbox{\titlehead}{
\begin{flushleft}
hep-th/9305010 \\
IC/93/83\\
May 1993\\
\end{flushleft}
}
\end{flushright}
\begin{center}
{\LARGE\bf Derivation of the Verlinde Formula}\\
\vskip .25in
{\LARGE\bf from Chern-Simons Theory}\\
\vskip .25in
{\LARGE\bf and the ${\bf G/G}$ Model}\\
\vskip .5in
{\bf Matthias Blau}\footnote{e-mail: blau@ictp.trieste.it}
and {\bf George Thompson}\footnote{e-mail: thompson@ictp.trieste.it}\\
\vskip .10in
ICTP\\
P.O. Box 586\\
I-34014 Trieste\\
Italy
\vskip .15in
\end{center}
\begin{abstract}
We give a derivation of the Verlinde formula for the $G_{k}$ WZW model from
Chern-Simons theory, without taking recourse to CFT, by calculating explicitly
the partition function $Z_{\Sigma\times S^{1}}$ of $\Sigma\times S^{1}$ with
an arbitrary number of labelled punctures. By a suitable gauge choice,
$Z_{\Sigma\times S^{1}}$ is reduced to the partition function of an Abelian
topological field theory on $\Sigma$ (a deformation of non-Abelian BF and
Yang-Mills theory) whose evaluation is straightforward. This relates the
Verlinde formula to the Ray-Singer torsion of $\Sigma\times S^{1}$.

We derive the $G_{k}/G_{k}$ model from Chern-Simons theory, proving their
equivalence, and give an alternative derivation of the Verlinde formula
by calculating the $G_{k}/G_{k}$ path integral via a functional version
of the Weyl integral formula. From this point of view the Verlinde formula
arises from the corresponding Jacobian, the Weyl determinant. Also,
a novel derivation of the shift $k\ra k+h$ is given, based on the index of
the twisted Dolbeault complex.
\end{abstract}
\end{titlepage}

\tableofcontents
\setcounter{footnote}{0}

\section{Introduction}

The Verlinde formula \cite{ver,ewcs,ms} is one of the most surprising and
interesting
results to have emerged from mathematical physics in recent years.
{}From the conformal field theory point of view it is a formula for
the number of conformal blocks of a rational conformal field theory
on a punctured Riemann surface. In this context it is a consequence
of the well-established, but nevertheless still somewhat enigmatic,
fact that the modular matrix $S$ implementing the modular transformation
$\tau \ra -1/\tau$ on the space of genus one conformal blocks
`diagonalizes' the fusion rules.

On the other hand, from the mathematical point of view, the Verlinde
formula is an expression for the dimension of the space of holomorphic
sections of some line bundle on a moduli space of vector bundles. As
such, it should yield to a standard mathematically rigorous derivation
but has withstood these attempts so far in all but the simplest of cases,
the $SU(2)$ Wess-Zumino-Witten (WZW) model. For some work in this direction
see e.g.~\cite{nnver}.\footnote{We understand that
a rigorous general proof has recently been obtained by Faltings as well
as by Narasimhan and Ramadas.}
In that case, denoting the
vector space of conformal blocks of the WZW model at level $k$ on a
genus $g$ surface $\S_{g}$ by $V_{g,k}$, the Verlinde formula reads
\be
\dim V_{g,k} = (\trac{k+2}{2})^{g-1}\sum_{j=0}^{k}\left[\sin\trac{
(j+1)\pi}{k+2}\right]^{2-2g} \;\;.\label{i1}
\ee
The expression on the right hand side of (\ref{i1}) has several notable
(and non-obvious) features, not the least of which are that it is indeed
an integer and a finite polynomial in $k$. Its generalization to arbitrary
compact Lie groups $\bG$ is
\be
\dim V_{g,k} = \left(C(k+h)^{r}\right)^{g-1}\sum_{\la\in\Lambda_{k}}
\prod_{\a\in\Delta}(1-e^{i\frac{\a(\la + \rho)}{k+h}})^{1-g}\;\;,\label{i3}
\ee
where $C,h,r$ are the order of the center, the dual Coxeter number, and the
rank of $\bG$, $\Lambda_{k}$ denotes the space of integrable highest weights
at level $k$, $\Delta$ the set of roots of $\bG$, and $\rho$ the Weyl vector
(half the sum of the positive roots). We want to draw attention to the fact
that in both these formulae the square of the Weyl denominator makes a
prominent appearance.

The purpose of this paper is to shed some light on these formulae (and
their counterparts for surfaces with labelled punctures) by deriving them
directly from Chern-Simons theory using only gauge theory techniques
and without taking recourse to either conformal field theory or mathematics
more advanced than elementary group theory. As the basic chain of
arguments leading
to the derivation of (\ref{i1},\ref{i3}) from Chern-Simons theory is
quite simple, we present it at the end of the Introduction. The length
of this paper is due to the fact that we have attempted to make it
reasonably self-contained and not to appeal to calculations which have
been performed elsewhere in superficially similar but different contexts
(here we have in mind in particular the derivation of the `shift' $k\ra
k+h$ of section 6 or the explicit evaluation of the Ray-Singer torsion
of $S^{1}$ and $\SS$ in section 3).

Our derivation, which is based on an `Abelianization' of the
Chern-Simons path integral via an appropriate gauge choice,
explains and highlights certain features of the Verlinde
formula and its relation to various topological field theories in two and
three dimensions. For instance, it explains the appearance of the square
of the Weyl denominator in the Verlinde formula by relating it to either
the Ray-Singer torsion of $\SS$ (section 3) or the Weyl integral formula
(section 5). Moreover, in the course of the derivation we also prove the
equivalence of Chern-Simons theory on $\SS$ with the $G/G$ gauged WZW model
on $\S$, thus establishing conjectures of Spiegelglas \cite{sp}
anticipated by Verlinde and Verlinde \cite{verver} and
partially verified  by Witten \cite{ewwzw}. We also establish the
equivalence of the $G/G$ model with a compact (`q-deformed') version
of BF theory which, like Yang-Mills theory, is related to {\em
classical} representation theory and has been extensively studied in
e.g.~\cite{btbf2,btym,ewym}. Via the methods of \cite{ew2d}
this brings one closer to a fixed point theorem
interpretation of the Verlinde formula (either for path integrals,
as we will explain in \cite{btfp}, or for some finite dimensional integral).
Finally, our derivation of the fusion rules
in section 7.6 shows that they arise naturally from Chern-Simons theory, and
already in diagonalized form, the traces of Wilson loops only ever
receiving contributions from those gauge field configurations where the
classical characters satisfy the quantum fusion rules.

The methods we use or develop in this paper can also be applied to
other problems arising e.g.~in coset models or Yang-Mills theory.
For instance, they lead to a very simple derivation
of the $2d$ Yang-Mills partition function (see \cite{btfp}).
They also provide us with Lagrangian realizations of $(G/H)/(G/H)$
models (e.g.~`$S^{2}/S^{2}$'-models) whose partition functions would
calculate the number of conformal blocks of the $G/H$ coset model.
These methods can also be applied to non-compact groups
like $SL(2,{\bf R})$ (where the conjugation into standard form needs
to be performed seperately for the elliptic, parabolic, and
hyperbolic elements).
In principle, they also allow us calculate Witten three-manifold
invariants of mapping tori $\S_{g}\times_{f} S^{1}$ directly from gauge
theory. It remains to be seen, however, if this can be turned into an
effective calculational tool for $g>1$.

\subsection{Some Background}

To explain why a derivation of the Verlinde formula from Chern-Simons theory
is possible in principle, and to indicate
how we will proceed in practice, we will have
to briefly recall some basic facts of Chern-Simons theory. For all the
other things that should be said about Chern-Simons theory,
see e.g.~\cite{ewcs,emss,axcs,pr}.

Choosing a closed oriented three-manifold $M$ and a compact gauge
group $\bG$ (which we will assume to be simply-connected so that
any principal $\bG$-bundle over $M$ is trivial), the Chern-Simons
action for $\bG$ gauge fields $\unA$ on $M$ (we reserve the notation $A$
for spatial gauge fields) is defined by
\be
kS_{CS}(\unA) =\trac{k}{4\pi}\int_{M}\tr(\unA d\unA + \trac{2}{3}\unA^{3})\;\;.
\ee
The trace (invariant form on the Lie algebra $\lg$ of $\bG$)
is normalized in such a way that invariance of $\exp ik S_{CS}$ under
large gauge transformations requires $k\in{\bf Z}$.

Traditionally, two approaches have been used to evaluate the partition
function of this action. One of them is perturbative in nature and is
conveniently performed in a background field expansion. In this theory
this amounts to expanding about flat connections. Then, to one
loop order, one finds that the effective theory is
\be
Z_{M}(S_{CS},k)^{(1)}(\unA_{c}) =
T_{M}(\unA_{c})^{1/2}\exp{\left( i(k+h) S_{CS}(\unA_{c}) \right)} \, ,
\label{1loop}
\ee
(modulo contributions coming from the ghost and one-form zero modes)
with the integration over the moduli space of flat connections $\unA_{c}$
still to be done. Here $T_{M}$ is the Ray-Singer torsion \cite{rs},
a particular
metric independent ratio of determinants of twisted Laplacians on $M$.

This is a Gaussian
approximation and on a general three manifold represents the first term
in a $1/k$ expansion of the path integral. Such a
perturbative approach is
particularly useful for exploring the relationship of Chern-Simons theory
with knot-invariants. At higher loops, however, it manages to hide quite
effectively the basic simplicity and elegance of Chern-Simons theory.

On the other hand, the relation of \CS\ theory to conformal field theory
can be used to evaluate the partition function non-perturbatively using
either surgery or Heegard splitting techniques,
see e.g.~\cite{ewcs,nncs}. In certain cases the large $k$ limit
of these results has been shown to compare favourably with the
evaluation of (\ref{1loop}).
What we will show in this paper is that for three-manifolds
of the form $\SS$ the partition function can be evaluated
exactly by ordinary gauge theory techniques (essentially because on these
manifolds there is a gauge choice available which makes the theory one-loop
exact).

An alternative approach to the quantization of \CS\ theory is canonical
quantization.
On three-manifolds of the form $\S\times{\bf R}$, \CS\ theory can be
subjected to a canonical analysis. Upon choosing the gauge $A_{0}=0$,
one determines the classical reduced phase space to be the moduli space
$\M$ of flat connections on $\S$. This is a symplectic space (as it should
be) and becomes K\"ahler once one chooses a complex structure on $\S$.
Hence one can use the recipe of geometric quantization to quantize this
system, the Hilbert space being the space of holomorphic sections of a
line bundle over $\M$ whose curvature is ($i$ times) the symplectic form.
It follows from Quillen's calculation \cite{qdet} and the fact that
the symplectic form for `level $k$' \CS\ theory is $k$ times the fundamental
symplectic form $\frac{1}{2\pi}\int_{\S}\d A \d A$, that the line bundle
in question is (for $SU(n)$) the $k$-th power of the determinant line bundle
associated to the family of operators $\{\bar{\del}_{A}\}$. In
\cite{ewcs} and \cite{verver} it is shown that
the space of holomorphic sections of this line bundle (the space of
states of \CS\ theory) coincides with the space $V_{g,k}$
of holomorphic blocks of the $\bG_{k}$ WZW model.

What is important for us is the fact that the dimension of this vector
space is given by a path integral of \CS\ theory. In fact, since the
Hamiltonian of \CS\ theory is zero (like that of any generally covariant
or topological theory), the statistical mechanics formula
\be
Z_{\SS} = \Tr e^{-\beta H}
\ee
for a circle of radius (imaginary time) $\beta$ reduces to
\be
Z_{\S_{g}\times S^{1}}(S_{CS},k) = \dim V_{g,k}\;\;.
\ee
This is the key equation which allows us to
derive the Verlinde formula (\ref{i1},\ref{i3}) by evaluating the
Chern-Simons partition function of $\S_{g}\times S^{1}$.
If one does not want to make
use of the identification of \CS\ states with conformal blocks, one
can nevertheless regard this derivation as a calculation of the dimension
of the \CS\ Hilbert space on $\S_{g}\times{\bf R}$.

\subsection{Outline of the Derivation}

The derivation of the Verlinde formula from Chern-Simons theory is
more or less straightforward but somewhat lengthy once one pays
due attention to certain technicalities. In order not to distract from
the basic simplicity of the argument,
in the following we sketch the main ideas and give a general outline of
the derivation. The presentation in this section will be informal.

The strategy will be to exploit the large gauge symmetry present in \CS\ theory
in a way which a) is compatible with the geometry of the problem, and b)
simplifies the action to the extent that the path integral can indeed be
evaluated explicitly.

The first of these desiderata we fulfill by choosing the gauge
$\del_{0}A_{0} = 0$ (the more obvious choice $A_{0}=0$ not being available
as $A_{0}$ may have a non-trivial holonomy along the circle). This
still leaves us with the time-independent gauge transformations
(and certain `large' time-dependent gauge transformations) to play with.
The former can be used to conjugate $A_{0}$ into the Lie algebra $\lt$ of
a maximal torus $\bT$ of $\bG$. By integrating over the time-dependent
modes of the $\lt$-components of $A$ and all the modes of the remaining
components of the gauge fields and the ghosts (all these integrals are
Gaussian) one is left with an effective two-dimensional {\em Abelian}
\tft\ (sections 2 and 3), which can then be easily evaluated.

Alternatively, one may wish to
trade the constant mode of $A_{0}$ for the holonomy $g = \exp A_{0}$
(section 4). If
one does that and eliminates the time-dependent modes of the gauge fields,
e.g.~by thinking of the path integral on $\SS$ as the trace of an
amplitude on the cylinder $\SI$ with (time-independent) boundary conditions,
one arrives at the action of the $G/G$ coset model. The fact that all
the determinants work out to produce precisely the Haar measure on the
group from the linear measure on the Lie algebra is one of the numerous
miracles we witness in this theory. As the notation
`$/G$' indicates, this model has a local $\bG$ gauge invariance which
can be used to conjugate $g$ into $\bT$ (section 5).
Again, the integration over
the non-Abelian components of $A$ is easily performed, leaving one
with the same effective two-dimensional Abelian theory.

In whichever way one proceeds one finds that the partition function
has now been reduced to the manageable form
\be
Z_{\SS}(S_{CS}) = \int D[\phi,A] \exp(i(k+h)\,S_{\phi F}(\phi,A))\;\;,
\ee
where $A$ is a $\lt$-valued gauge field and
$\phi$ is a compact scalar field taking values in $\bT$. The
action (a compact Abelian BF action) and the measure are
\be
S_{\phi F}(\phi,A)=\trac{1}{2\pi}\int_{\S}\tr \phi F_{A}
\ee
and
\be
D[\phi,A] = D\phi\,DA\,\det({\bf 1}-{\rm Ad}(e^{i\phi}))^{1-g}
\ee
respectively.
The action is rather obviously a remnant of the \CS\ action. What deserves
attention, though, are the occurrence of the shift $k\ra k+h$ and the
non-trivial measure. The former could have been more or less anticipated
from other investigations of \CS\ theory. Here it arises from a gauge invariant
regularization of the ghost and gauge field determinants (via the
index of the twisted Dolbeault complex, section 6).

These determinants, which
also almost cancel each other (a typical feature of topological field
theories), leave behind the finite dimensional
determinant appearing in the measure.
{}From the point of view of sections
2 and 3 it arises as the square-root of the Ray-Singer torsion of $\S_{g}
\times S^{1}$.
In fact, by a theorem of Ray and Singer one has $T_{\S\times S^{1}}
= T_{S^{1}}^{2-2g}$,
while the torsion of a gauge field with holonomy $t\in\bT$ on the
circle is just $\det({\bf 1} - {\rm Ad}(t))$ (section 3.1), so that the
usual one-loop approximation (\ref{1loop})
would suggest the appearance of such a term.
We want to emphasize, however, that the above result is exact.

{}From the $G/G$ model point of view, on the other hand, the measure
can be understood as the Jacobian arising in the Weyl integral formula
(section 5.1) relating the integral over $\bG$ of a conjugation invariant
function to an integral over $\bT$ (and almost cancelled by the determinant
arising from the gauge field integration).

The above Abelian $\phi$F theory is closely related to ordinary
non-Abelian BF theory with action $\int\tr B F_{A}$, $B$ a non-compact
Lie algebra valued scalar field, and Yang-Mills theory (whose action
differs from that of BF theory by a term $\sim B^{2}$). In both
these theories the partition function can be expressed as a sum over
all irreducible representations of $\bG$ \cite{btym,ewym}. Here the
compactness of $\phi$ imposes a cutoff on the representations contributing
and reduces the integral to a sum over integrable representations.
To see this directly,
one can make a change of variables from $A$ to its curvature $F_{A}$
(plus longitudinal components). The $F_{A}$-integral (constrained
by the condition that $F_{A}$ be an integral two-form) will then
impose a delta function constraint on $\phi$ allowing only certain discrete
values of $\f$ (corresponding to the highest weights of the integrable
representations) to contribute.
The details of this calculation
will be explained in section 7, where we also discuss the inclusion of
punctures (`vertical' Wilson loops in \CS\ theory).
The result is then precisely the
Verlinde formula, up to an overall normalization which we determine in
section 7.5 (again taking care not to make use of results from
conformal field theory or the representation theory of loop groups).

\section{Chern-Simons Theory on ${\bf \Sigma \times S^{1}}$}

In the Introduction we mentioned the usual approach to the
evaluation of the \CS\ partition function via a background field
expansion.
On a three manifold of the form $\Sigma \times S^{1}$ one may follow a
different approach. This
will give us an exact evaluation of the path integrals involved but,
nevertheless, has some features in common with the background field
method. At an intermediate stage of the calculation we will obtain a
formula quite like (\ref{1loop}), except that only gauge fields with
values in the Cartan subalgebra $\bft \subset \lg$ of the Lie algebra
enter and, indeed, they are not flat (for what is meant by the
Ray-Singer torsion in this case see section 6.2).
The final step in the calculation is then
easy to perform. The reason that one may go so far in this instance is
that for a manifold of the type $\Sigma \times S^{1}$ there is a gauge
choice available which ensures that all of the path integrals that we
will encounter are Gaussian.

\subsection{Gauge Fixing}

\noindent It makes sense, on $\Sigma \times S^{1}$, to split the gauge field
into components
\be
\unA =   A + A_{0}dt \;\; ,
\ee
in an obvious notation. In order to perform the path integral we will need
to fix on a gauge. On
the line, with suitable boundary conditions, it would be possible to set
$A_{0} =0$. On the circle, however, this gauge choice is not
possible. There is a simple reason for this. In terms of Fourier modes,
the gauge transform of the gauge field acquires the shift $n\Lambda_{n}
+ \ldots$,
so that all the non-constant modes may be eliminated by an appropriate
choice of the $\Lambda_{n}$. This shift is absent in the case of the
constant mode. This discussion shows us that the natural choice here is
the gauge
\be
\del_{0} A_{0} =0 \;\; . \label{gf1}
\ee
This eliminates almost all of the time dependent gauge transformations,
see (\ref{lgt}).

But time independent transformations can still
be used to impose a stronger restriction on $A_{0}$, and it is a
particular such choice which will simplify matters to the extent that
the Chern-Simons partition function becomes explicitly calculable.
In order to motivate this choice of gauge condition
it is useful to pass to a more group theoretic description of the above
discussion.

We denote by $g$ the holonomy of $A_{0}$ around the circle $S^{1}$,
i.e.~$g$ is the path ordered exponential
\be
g=P\exp\left(\oint A_{0} \right)\;\;.
\ee
Under a gauge transformation by $h$, the holonomy $g$ is conjugated,
\be
A_{0}\ra A_{0}^{h}\equiv h^{-1}A_{0}\,h+h^{-1}\del_{0}h\Rightarrow
g\ra h^{-1}g h\;\;.
\ee
This also makes it clear that one may not impose the gauge $A_{0}=0$,
for that would mean that we could conjugate $g$ to the identity,
which is not possible unless $g$ already happens to be the identity.
What we may do, however, is conjugate any $g \in \bG$ into a maximal torus
$\bT$ of the group $\bG$, i.e.~we can `diagonalize' $g$,
and we learn that the only gauge invariant
degree of freedom of a gauge field on the circle is the conjugacy
class of its holonomy. Thus, if we eliminated $A_{0}$ in favour
of $g$, it would be possible to impose the gauge condition $g\in\bT$.
This is the procedure we will adopt in section 5 to abelianize the
$G/G$ action. Here, however, we are still working at the Lie algebra level,
and we will instead make use of the possibility to conjugate an
element of the Lie algebra $\lg$ into the Lie algebra $\lt$ of $\bT$.
We may thus augment (\ref{gf1}) with the condition
\be
A_{0}^{\lk}=0 \;\;, \label{gf2}
\ee
where we have split the gauge field into the part $A^{\lt}$ taking
values in $\lt$ and into $A^{\lk}$
taking values in the complement $\lk$ of $\lt$ in $\lg=\lt\oplus\lk$.

In order to impose (\ref{gf1}) and (\ref{gf2}) simultaneously we add the
following gauge fixing terms to the action,
\be
\int_{\Sigma \times S^{1}} \tr \left[ BA_{0} + \bar{c}D_{0} c \right]
\, , \label{gfa}
\ee
where
\be
D_{0}= \del_{0} + A_{0} \;\;,
\ee
and we furthermore impose the conditions
\be
\oint B^{\lt} = \oint c^{\lt} =\oint \bar{c}^{\lt}=0 \;\; . \label{cond}
\ee
This may need some explanation. Firstly, (\ref{gfa}) is BRST
exact and so does not generate any unwanted metric dependence in the
path integral. The conditions (\ref{cond}) may also be imposed in a
BRST invariant manner, and so also do not give rise to spurious metric
dependence (the way to do this is explained e.g.~in \cite{btbf2,pr}).
For this reason we
have refrained from introducing the metric explicitly in the above
expressions. The condition on the
multiplier field $B$ is clearly needed so as to impose precisely the
conditions (\ref{gf1}) and (\ref{gf2}). The condition on the anti-ghost
$\bar{c}$ follows from the requirement that the anti-ghost modes be in
one to one correspondence with those of the multiplier field.

Perhaps it is surprising
that there is also a constraint on the ghost $c$. But that this
requirement is correct is easily seen if we note that the combined conditions
(\ref{gf1}, \ref{gf2}) are invariant under constant torus valued gauge
transformations. These transformations are therefore not used to arrive
at the gauge fixing conditions that we have chosen and hence those
components of the ghost should not appear.

\subsection{One Loop Exactness}

\noindent We wish to perform the path integral over all the modes of
$A^{\lk}$
and the ghosts and the non-constant (in time)
modes of $A^{\lt}$. This will leave us
with an effective two-dimensional Abelian theory in $A_{0}^{\lt}$ and
(the constant mode of)
$A^{\lt}$. The modes to be integrated over enter
quadratically in the action so the integrals to be performed are simply
Gaussian. Hence, as far as all these modes are concerned, the one-loop
approximation is exact. The Chern-Simons path integral then becomes a two
dimensional path integral. The reason for approaching the problem in
this way is that at this point we may employ techniques developed for
two-dimensional gauge theory to complete the evaluation of the
three-dimensional path integral.

In the gauge chosen the (quantum) Chern-Simons action takes on a
particularly simple form,
\be
S_{CS}(\unA)=
\int_{\Sigma \times S^{1}}\! dt\,\tr \left( A_{0}^{\lt}d A^{\lt} +
A^{\lt} \del_{0}A^{\lt} +
A^{\lk}D_{0}A^{\lk} + \bar{c}^{\lk}D_{0}c^{\lk} +
\bar{c}^{\lt}\del_{0}c^{\lt} \right) \;\; , \label{act1}
\ee
where it is understood that $A_{0}$, $c$ and $\bar{c}$ satisfy the
constraints (\ref{gf1}), (\ref{gf2}) and (\ref{cond}).

Performing the integration over the modes specified above generates the
following ratios of determinants:
\be
\frac{\Det_{\bft}' [\del_{0}]_{\Omega^{0}(\Sigma) \otimes  \Omega^{0}(S^{1})}}{
\Det_{\bft }'^{1/2}
[\del_{0}]_{\Omega^{1}(\Sigma) \otimes
\Omega^{0}(S^{1})} } \, . \, \frac{\Det_{\bfk} [D_{0}]_{\Omega^{0}(\Sigma)
\otimes  \Omega^{0}(S^{1})}}{
\Det_{\bfk }^{1/2}
[D_{0}]_{\Omega^{1}(\Sigma) \otimes
\Omega^{0}(S^{1})} } \;\;. \label{ratio}
\ee
The subscript labels refer to the spaces on which the
operators act, and the prime
indicates that one should not include the $S^{1}$ zero mode.
We will perform a precise analysis of this ratio of determinants in
section 6. Some insight into what the final result will look like,
and why, can however also be gained by proceeding in a less rigorous
fashion and this is what we will do here.

The first thing to note is that the determinants almost cancel, as a
one-form in two dimensions
is `like' a pair of scalars. To understand the qualification `almost'
in the above, recall that by the Hodge decomposition theorem
any $p$-form $\omega$ on a compact $n$-dimensional Riemannian manifold $M$
may be uniquely written as the sum of an exact, a coexact, and a harmonic
form,
\be
\omega = d\alpha + d^{*}\beta + \gamma\;\;.\label{hodge}
\ee
Here $d^{*}$
is the adjoint of $d$ with respect to the scalar product
\be
(\omega_{1},\omega_{2}) = \int_{M}\omega_{1}*\omega_{2}\;\;.\label{scal}
\ee
It follows from $*^{2}=(-1)^{p(n-p)}$ that, acting on $p$-forms,
\be
d^{*}=(-1)^{np + n + 1}*d*\;\;.
\ee
Thus, in two dimensions, (\ref{hodge}) refines the above statement
in the sense that it expresses a one-form in terms of two scalars
$\alpha$ and $*\beta$ (modulo constants which are the harmonic zero forms)
and a harmonic form representing an element of $H^{1}(\S,{\bf R})$.
This allows us to figuratively decompose the space of one-forms as
\be
\Omega^{1}(\Sigma) = \Omega^{0}(\Sigma) \oplus \Omega^{0}(\Sigma) \oplus
H^{1}(\Sigma) \ominus 2H^{0}(\Sigma)\;\; .
\ee
Recalling that the harmonic modes are orthogonal to the others we may
then deduce that (\ref{ratio}) is more or less equal to
\be
\frac{\Det_{\bft}' [\del_{0}]_{H^{0}(\Sigma) \otimes  \Omega^{0}(S^{1})}}{
\Det_{\bft }'^{1/2}
[\del_{0}]_{H^{1}(\Sigma) \otimes
\Omega^{0}(S^{1})} } \, . \, \frac{\Det_{\bfk} [D_{0}]_{H^{0}(\Sigma)
\otimes  \Omega^{0}(S^{1})}}{
\Det_{\bfk }^{1/2}
[D_{0}]_{H^{1}(\Sigma) \otimes
\Omega^{0}(S^{1})} } \, . \label{ratio1}
\ee
This mechanism for the cancellation of modes is similar to that
used by Witten in \cite{ew2d}. However, (\ref{ratio1})
is not quite right on two counts. Firstly, as $A_{0}$ has both
harmonic and non-harmonic modes it mixes the two in the determinants.
Secondly, to give
meaning to these functional determinants one should regularise. Our
expectation is that a gauge invariant regularization scheme will
reproduce (\ref{ratio1}) plus the shift $k \rightarrow k + h$.
As the latter is not our most immediate concern, we postpone a
discussion of this issue until section 6. There we will also discuss
the case of non-constant $A_{0}$. For our present purposes, however, it
is enough to take $A_{0}$ to be a constant (and hence flat) as the
integral over $A^{\lt}$ will eventually impose a delta function
constraint to this effect. With $A_{0}$ understood to be constant the
determinants become
\bea
T_{S^{1}}(A_{0})^{\chi(\Sigma)/2} & \equiv &
\left[\Det_{{\bf t}}'(\del_{0})_{\Omega^{0}(S^{1})}\, . \,\Det_{{\bf
k}}(D_{0})_{\Omega^{0}(S^{1})} \right]^{b_{0}-b_{1}/2} \nonumber \\
&=&  \left[\Det_{{\bf
t}}'(\del_{0})_{\Omega^{0}(S^{1})}\, . \, \Det_{{\bf
k}}(D_{0})_{\Omega^{0}(S^{1})}  \right]^{\chi(\Sigma)/2} \label{dets}
\eea
where the Betti numbers $b^{i} = {\rm dim} H^{i}(\Sigma)$ are
$b^{0}=1, b^{1}=2g$.
In the next section we show that (\ref{dets}) equals the
square root of the Ray-Singer analytic torsion of $\SS$
(and that $T_{S^{1}}(A_{0})$ is the torsion of the circle,
explaining the notation adopted in (\ref{dets})).

As all the non-constant time modes have been integrated out, the
Chern-Simons path integral on $\Sigma \times S^{1}$ becomes a path integral
on $\Sigma$. This path integral is
\be
\int DA_{0\,}DA^{\lt} \, T_{S^{1}}(A_{0})^{\chi(\Sigma)/2} \, \exp{
\left(\trac{ik}{2\pi} \int_{\Sigma} A_{0} F_{A} \right)} \, .\label{zcs}
\ee

As this bears a formal resemblance to the result (\ref{1loop})
of a one-loop calculation in a background field, it is useful
to point out the differences as well. First of all, we have not
chosen a background field about which to expand. Secondly, we
have therefore also not made any approximation (i.e.~dropped
higher orders in the quantum field). And finally, to arrive at
(\ref{zcs}) we have only integrated over parts of the modes,
whereas the one-loop background field expansion entails an integral
over all quantum fields.

\subsection{A Supersymmetry}

\noindent The cancellation of contributions between the one-forms and
zero-forms may be formalised by introducing a supersymmetry relating the
two. Indeed (\ref{act1}) has a number of BRST like symmetries, the one
of interest to us here being
\be
\d A^{\lk}= \eps c^{\lk} \, , \;\;\;\;\;\;
\d \bar{c}^{\lk}= 2 \eps A^{\lk} \,, \label{sup}
\ee
where $\eps$ is a one form on $\Sigma$. Now (\ref{sup}) allows us to
conclude that the non-harmonic modes of $A^{\lk}$ are precisely paired
against the non-harmonic modes of $c^{\lk}$ and of $\bar{c}^{\lk}$.

For the harmonic modes the situation is somewhat different. In this case
we need to take $\eps$ to be harmonic. The first of (\ref{sup}) shows us
that, as there is only one $c^{\lk}$ harmonic mode, only one of the $g$
$A^{\lk}$ zero-modes is paired against the zero-mode of $c^{\lk}$. The
other $g-1$ modes are not transformed in (\ref{sup}). The
second of the equations (\ref{sup}) can be satisfied for some
$\bar{c}^{\lk}$, when $\eps$ is harmonic and the gauge field is taken to
be the harmonic mode that makes an appearance in the first transformation.

There is a similar supersymmetry
relating the non-constant modes of $A^{\lt}$ to those of $c^{\lt}$ and
$\bar{c}^{\lt}$. These considerations lead us once more to the conclusion
that the ratio of determinants that appears in (\ref{ratio1}) is given
correctly by (\ref{dets}).

We will also find a similar supersymmetry in
the (appropriately gauge fixed) $G/G$ model, see section 6.
There we will also give a more rigorous derivation of the
above result. In particular, using a heat kernel regularization,
the remaining
finite dimensional determinant will be seen to arise as a consequence
of the gravitational contribution to the chiral anomaly.

\section{Ray-Singer Torsion}

We will show presently that the ratio of determinants that we have to
evaluate in (\ref{dets}) is known as
the Ray-Singer Torsion for the circle. Formally, given a flat vector
bundle with flat connection $A$
over a Riemannian manifold $M$ of dimension $n$ and denoting the
corresponding twisted Laplacian on $k$-forms by
\be
\Delta_{k}= d_{A}d_{A}^{*}+d_{A}^{*}d_{A} \label{lap}
\ee
the Ray-Singer Torsion \cite{rs}
is defined to be\footnote{There is some latitude in the definition of the
Torsion. Here we have adopted the definition of Ray and Singer, though
the reader should be warned that some authors find it more convenient
to call the inverse of this the Torsion.}
\be
T_{M}(A) =
\prod_{k=0}^{n} [\Det{\Delta_{k}}]^{(-1)^{k+1}k/2} \;\;, \label{rst1}
\ee
provided that the $\Delta_{k}$ have no harmonic modes (i.e.~the twisted
de Rham complex is acyclic). Even in that case, however,
the product, as it stands, is still not well defined. To give meaning to
the infinite product of eigenvalues, Ray and Singer introduced the
$\zeta$-function regularization of determinants. Each determinant that
appears in (\ref{rst1}) is defined by
\be
\Det{\Delta} = \exp{\left( -\zeta'(0) \right)} \, ,
\ee
with
\be
\zeta(s) = \frac{1}{\Gamma(s)}\int_{0}^{\infty} t^{s-1} \tr
\exp(-t\Delta)dt \;\;.
\ee
The remarkable fact established by Ray and Singer
is that, with these definitions, the Ray-Singer Torsion does not depend on
the Riemannian metric that went into its definition.

If there are non-trivial cohomology groups $H^{k}_{A}$,
there are two things that need to be
changed in the above. First of all, the $\zeta$-function will have to
include a projector onto the non-zero eigenvalues of $\Delta_{k}$.
Technically this is achieved by defining
\be
\zeta(s) = \frac{1}{\Gamma(s)}\int_{0}^{\infty} t^{s-1} \tr
\left(\exp{(-t\Delta)} -P\right) dt \, ,
\ee
where $P = \lim_{t \rightarrow \infty} \exp{(-t\Delta)}$ is the projector
onto the harmonic modes. Secondly, the Ray-Singer torsion should then
properly be thought of as an object assigning a number to a choice of
volume element on the cohomology groups, i.e.~as an element of the
one-dimensional vector space
\be
T_{M}(A)\in\bigotimes_{k=0}^{n}(\det H^{k}_{A})^{(-1)^{k}}\;\;.
\ee
Here $\det V$ denotes the highest exterior power of $V$,
\[\det V = \bigwedge^{{\rm dim}\,V}V\;\;.\]
With the appropriate definitions the torsion can then again be shown to
be metric independent. This minor ambiguity (a scale factor) in the
definition of the Torsion $T_{M}(A)$ as a number will not concern us in the
following as we will use a different kind of argument to fix the overall
normalization of the path integral.

In a further development Schwarz \cite{asrs} gave simple field theoretic
representations of the Ray-Singer Torsion. This form of the Torsion
allows for standard path integral manipulations and so makes its
determination, in good circumstances, possible. We use such a
representation to make contact with Chern-Simons theory.
{}From the path integral point of view, the above ambiguity is also
easy to understand. It corresponds to the ambiguity one encounters
when attempting to gauge away or soak up the zero modes in the
path integral. For more
on the relation between the Ray-Singer torsion and path integrals,
see \cite{btbf2}.

\subsection{Ray-Singer Torsion on $S^{1}$}

\noindent
All gauge fields on a circle are flat so that it makes sense to define
the Torsion for any connection $A_{0}$
on $S^{1}$. From (\ref{rst1}) we see that, in this case,
\be
T_{S^{1}}(A_{0}) = [\Det{\Delta _{1}}]^{1/2} \;\;,
\ee
where the positive square root is to be taken.
It is possible to get rid of the troublesome square root since
\be
\Delta_{1} = d_{A_{0}}*d_{A_{0}}* = (d_{A_{0}}*)^{2}\;\;.
\ee
The last line makes sense as one-forms and zero-forms are in one to one
correspondence via the Hodge operator. We therefore want to calculate
$\det{d_{A_{0}}*}$. The field theoretic form for this determinant is
\be
T_{S^{1}}(A_{0}) = N \int_{\lg} D\e D\eb  \, \exp{\left(i\int_{0}^{1}\! dt\,
 \eb(t)(\del_{0} +A_{0})\e(t) \right)} \, , \label{rst}
\ee
with periodic boundary conditions on the anti-commuting fields $\e$ and
$\eb$. These
fields take values in the adjoint representation of ${\bf g}$ and the subscript
on the integral ($\lg$ or $\lt$ or $\lk$) will be used to indicate which of
those fields we are integrating over.
By making use
of the gauge invariance of $T_{S^{1}}(A_{0})$,
it is possible to give a simple evaluation of the path integral in
(\ref{rst}). From (\ref{rst}) we have that
\be
T_{S^{1}}(A_{0}^{g}) = T_{S^{1}}(A_{0}) \, . \label{rstg}
\ee
On the circle, as we have seen, any gauge field is gauge equivalent to a
constant gauge field (on the line it would be gauge equivalent to the
zero connection).
By making use of time independent gauge transformations the constant
gauge field may be conjugated into a given torus of the Lie algebra of the
group under consideration. By (\ref{rstg}) in order to evaluate the path
integral we need only consider the gauge field to be constant and to lie
in a torus. At this point we encounter a technical difficulty. The $\e$
and $\eb$ lying in the Cartan subalgebra have zero modes that do not
appear in the action so that at this point the path integral vanishes.
These are precisely the (twisted) harmonic modes that we have been
instructed to drop. We may simply define the right hand side of (\ref{rst})
to be an integral over the fields with values in $\bf k$.

In path integral language
this amounts to gauge fixing the zero modes to zero (there is clearly
enough symmetry to do this and is the analogue of projecting them out)
and to soaking up the contribution of the
rest of the modes lying in $\bft$ into the normalisation of $T_{S^{1}}(A_{0})$.
This factor is
\bea
\int_{\bft} D\e D\eb \exp{ \left( \oint \eb\del_{0} \e(t) \right) } & = &
\prod_{n>0} n^{2 \dim{\bft}}  \nonumber \\
&=& \exp{ \left( 2 \dim{\bft} \sum_{n>0} \ln{n} \right) } \nonumber \\
&=& 1/(2 \pi)^{\dim{\bft}}\label{tdet}
\eea

We have just evaluated one of the determinants that appear in
(\ref{dets}), namely
\be
\Det_{\lt}'(\del_{0})_{\Omega^{0}(S^{1})}\, = 1/(2 \pi)^{\dim{\bft}} \,.
\ee
The part of the path integral that is left to evaluate runs over the
fields with values in $\bfk$. This is well defined, up
to regularization, and corresponds precisely to the other determinant
that we encountered in Chern-Simons theory (\ref{dets}).
This functional integral is a standard representation of
\be
 \tr_{\bfk}[(-1)^{F}
 \exp{\left( {\rm ad}( A_{0}) \right) }] \, ,
\ee
with the trace restricted to $\bfk$. By linear algebra this is simply
\be
T(A_{0}) \sim \tr_{\bfk}[(-1)^{F} \exp({\rm ad} A_{0})]  =
\det[{\bf 1}-{\rm Ad}_{\lk}(\exp A_{0})] \, .\label{id}
\ee
In order to fix the constants and at the same time answer some questions about
regularization we compute the path integral.
One could do this by expanding $\e$ and $\eb$ in
Fourier modes but we follow a different path
here, which will prove useful when we wish to exhibit the equivalence of
the $G/G$ models and Chern-Simons theory on $\Sigma \times S^{1}$.

First define the path integral on the interval with boundary conditions
$\e(0) =\e$ and $\eb(1)= \eb$,
\be
Z[A_{0};\e,\eb] = \int_{\bfk} D\e D\eb  \, \exp{ \left( i\int_{0}^{1}\! dt\,
\eb(t)(\del_{0} +A_{0})
\e(t) -i\eb(1)\e(1) \right) } \, . \label{bdry}
\ee
The boundary term is needed to ensure that the $\e(t)$ variation of the
action is well defined. The path integral on the circle is now given by
\be
Z[A_{0}] = \frac{1}{(2 \pi)^{\dim{\bft}}} \int_{\lk}
d\e d\eb e^{i\eb \e} Z[A_{0};\e,\eb] \, .
\ee
In order to evaluate (\ref{bdry}) we perform a gauge
transformation
\be
\e(t) \rightarrow {\rm Ad}(g_{t})\e(t) \, , \, \eb(t) \rightarrow
{\rm Ad}(g_{t})\eb(t) \, ,
\ee
with the gauge parameter defined by
\be
g_{t} = P \exp{ \left( \int_{0}^{t}A_{0} \right) } =  \exp{ \left( tA_{0}
\right)
} \, .
\ee
Clearly $g_{0}=1$ and we set $g_{1} = \exp A_{0} \equiv  g$.
With respect to the new fields (\ref{bdry}) becomes
\be
\int_{\bfk} D\e D\eb  \, \exp{ \left( i\int_{0}^{1} dt \eb(t)\del_{0}
\e(t) -i{\rm Ad}(g^{-1})\eb \e(1) \right) } \, ,
\ee
the Jacobian of the transformation being unity. The only point to note
is that in terms of the new fields the boundary data does not change for
$\e(t)$, $\e(0)=\e$, while for $\eb(t)$ one has $\eb(1)=
{\rm Ad}(g^{-1})\eb$. This means that
\be
Z[A_{0};\e,{\rm Ad}(g)\eb] =  \int_{\bfk} D\e D\eb  \, \exp{ \left(
i\int_{0}^{1}
dt \eb(t)
\del_{0} \e(t) -i\eb \e(1) \right)} \, , \label{bdry1}
\ee
with $\e(0)=\e$ and $\eb(1)=\eb$. We may now evaluate (\ref{bdry1})
by shifting fields
\bea
\e(t) & = & \e + \e_{q}(t) \, , \nonumber \\
\eb(t) &=& \eb + \eb_{q}(t) \, ,
\eea
so that the boundary conditions on the ``quantum'' fields are $\e_{q}(0)
=0$ and $\eb_{q}(1)=0$. This leads to
\be
Z[A_{0};\e,{\rm Ad}(g)\eb] =
\exp{\left( -i\eb \e \right)} \,\int_{\bfk} D\e_{q} D\eb_{q}
 \, \exp{ \left( i\int_{0}^{1} dt \eb_{q}(t)
\del_{0} \e_{q}(t) \right) } \, .
\ee
This path integral
may now be thought of as a path integral on the circle,
with the zero mode neglected. It can be evaluated just as in (\ref{tdet})
to be
\be
\int_{\bfk} D\e_{q} D\eb_{q}
 \, \exp{ \left( i\int_{0}^{1} dt \eb_{q}(t)
\del_{0} \e_{q}(t) \right) }=
1/(2\pi)^{\dim\lk}\;\;.
\ee
Putting all the pieces together we find, as expected from (\ref{id}),
\bea
Z[A_{0}] &=& \trac{1}{(2\pi)^{\dim{\lg}}}
\int d\e d\eb \, \exp{ \left(i \eb (1  - {\rm Ad}(g^{-1})) \e \right) }
\nonumber \\
&=& \trac{1}{(2 \pi)^{\dim{\bfg}}}
\det{\left( {\bf 1}-{\rm Ad}_{\bfk}(g) \right)} \, .
\eea
In going to the second line of this equation we have made use of the fact
that $\det{\rm Ad}(g)=1$ (for the groups under consideration)
and that ${\rm dim}(\lk)$ is even.

\subsection{Normalisation of the Ray-Singer Torsion}

We would like to fix the constant $N$ that appears in (\ref{rst}). Recall
that for the trivial connection, when one ignores harmonic modes, the torsion
should be unity.
As this amounts to projecting out the zero modes
of $\e$ and $\eb$ in this instance, we have
\be
T_{S^{1}}(A_{0}=0) = N \Det'\del_{0} = N/(2\pi)^{\dim\lg} \stackrel{!}{=}
1\;\;,
\ee
so that
\be
T_{S^{1}}(A_{0}) = N Z[A_{0}] = \det({\bf 1} - {\rm Ad}_{\lk}(g))\;\;.
\ee
This agrees with the results obtained by Freed \cite{dfcs} and
Witten \cite{ewym} using
spectral sequences and a Meyer-Vietoris argument.

\subsection{Ray-Singer Torsion on $\Sigma \times S^{1}$}

\noindent The discussion above shows us that the determinants encountered
in the Chern-Simons theory, for constant $A_{0}$, are representations of
the Ray-Singer Torsion on the circle and for that Torsion one has the
concrete form
\be
T_{S^{1}}(A_{0}) = \det{\left( {\bf 1}-{\rm Ad}_{\bfk}(g) \right)} \, .
\ee
Comparing with the semi-classical approximation (\ref{1loop}) we would
expect to obtain the square root of the
Ray-Singer Torsion $T_{\Sigma \times S^{1}}$ of
$\Sigma \times S^{1}$ rather than just some power of the Torsion on the
circle. That these are indeed the same follows from (a slight generalization
of) a theorem of Ray and Singer. This theorem \cite{rs} states that
if $M$ is a closed simply connected even-dimensional manifold,
then the torsion of the product manifold $M\times N$ (no restrictions
on the dimension or fundamental group of $N$) is
\be
T_{M \times N} = T_{N}^{\chi(M)} \, . \label{prod}
\ee
The theorem then applies to our case if $M = S^{2}$ and
$N=S^{1}$ and
\be
T_{S^{2}\times S^{1}}(A_{0}) = T_{S^{1}}(A_{0})^{2}
\ee
agrees with the result we have obtained. So what about higher genus
surfaces? Looking at the proof of the theorem one sees that
the restriction to simply connected $M$ is there to ensure
that a flat connection on $M \times N$ is completely
specified by its holonomies around the non-trivial $1$-cycles of $N$.
This will be the case for arbitrary flat vector bundles over $M \times
N$ provided that $M$ is simply connected.

Let us turn our attention to
those flat vector bundles over $M \times N$, where $M$ is not
necessarily simply connected, with connections of the local form
$A \in \Omega^{1}(N)\otimes \Omega^{0}(M)$, up to gauge equivalence. For
such bundles the connection is once more completely specified by its
holonomies around the $1$-cycles of $N$. In this setting (\ref{prod}) is
once more correct and is then a slight extension of the theorem of Ray
and Singer. As these conditions are met in the case at hand (after
all, we are calculating the Ray-Singer torsion of $A_{0}\,dt$ on $\SS$),
we find that
\be
T_{\SS}(A_{0}) = T_{S^{1}}(A_{0})^{\c(\S)}\;\;,
\ee
so that the determinant appearing in (\ref{zcs}) is indeed precisely
the square root of the torsion of $\SS$.

One may give a standard mathematical proof of this generalization following
almost line for line that of Ray and Singer. Turning the argument
on its head, we see that path integral manipulations can be used to
provide an alternative proof of this theorem, solving a problem raised
in \cite{btbf2}.

\subsection{Relationship with Yang-Mills Theory on $S^{1}$}

\noindent We could have expected the result we have derived for the
Ray-Singer Torsion on the circle on general grounds. Suppose we wished
to construct a Yang-Mills theory on the circle. Then, as $F_{A}$ is
automatically zero, the only terms that would appear in the action are
the gauge fixing and ghost terms. But the path integral obtained in this
way is simply (\ref{rst}) integrated over all gauge fixed connections,
with $\e$ and $\eb$ playing the role of the ghost $c$ and the
antighost $\bar{c}$ respectively. This
would be the integral of the Ray-Singer Torsion over the maximal Torus
of the group $\bG$ (or more precisely over $\bT/W$, where $W$ is the
Weyl group; see section 5.1 for the group theory used in the following).

Instead of working at the level of the connection, we could define
Yang-Mills theory directly in terms of the structure group. That is we
could simply associate to the connection its holonomy (this is like the
first step of gauge fixing, namely $\del_{0}A_{0}=0$).
The Yang-Mills path integral would then devolve to
\be
\int_{\bG} dg = 1 \, ,
\ee
with a particular choice of Haar measure. However, in our previous
considerations, we had moved down to the maximal Torus of $\bG$ (this
corresponds to the next stage of gauge fixing, $A_{0}^{\lk}=0$). We would
like, therefore, to express the integral over $\bG$ as an integral over
$\bT$. This may be done using the Weyl integral formula
(section 5.1) to give
\be
\int_{\bG} dg = \trac{1}{|W|} \int_{\bT}
\det{\left({\bf 1}- {\rm Ad}_{\bfk}(t)
\right)} dt \, ,
\ee
in which we recognise the Ray-Singer Torsion.

As an aside we would like to point out that this Yang-Mills theory
calculates a topological invariant in its own right. There is a theorem
that states that
\be
\chi(G/T) = \int_{\bT} \, \det({\bf 1} - {\rm Ad}_{\bfk}(t) ) dt \, ,
\ee
and this suggests that the properly normalised Yang-Mills path integral on
the circle is the Euler character of the homogeneous space $G/T$. It is
not too difficult to see that the gauge fixed Yang-Mills action is
equivalent to $N=2$ supersymmetric quantum mechanics on $G/T$, thus
guaranteeing that this path integral yields the Euler character.
This statement may be made more palatable by noting that the one-dimensional
Yang-Mills action is $Q$-exact ($Q$ is the BRST operator),
\be
S = Q\oint\tr \bar{c}^{\lk}A_{0}^{\lk}\;\;.
\ee
This is one of the characteristic features of cohomological field theories.
Furthermore there is a second supersymmetry obtained by exchanging $c$ and
$\bar{c}$, suggesting that this model is related to de Rham cohomology.
And finally these BRST symmetries are seen to be the typical
topological `shift' symmetries at the group level. Namely, for
\be
g(t) = P\exp(\int_{0}^{t}\! A_{0})g(0)
\ee
one finds
\be
Qg(t) = c(t)g(t)\;\;,
\ee
as one also transforms $g(0)$ according to $Qg(0)=c(0)g(0)$. This shows
that we have a Witten type supersymmetric quantum mechanics model, so
it is bound to calculate the Euler number of some space and the analysis
may proceed from here.

\subsection{The Path Integral to be Evaluated}

\noindent We have now reduced the three dimensional path integral to a
well specified two dimensional theory, namely
\be
\int DA_{0} DA^{\lt}
\det{\left( {\bf 1}-{\rm Ad}(e^{A_{0}}) \right )}^{\chi(\Sigma)/2}
\exp{ \left(\trac{ik}{2\pi} \int_{\Sigma} A_{0} F_{A} \right )} \, .
\ee
However, this is not quite the end of the story yet. It should be
kept in mind that the gauge conditions
\[ \del_{0}A_{0} = 0 \;\;,\;\;\;\;\;\;A_{0}^{\lk}=0\;\;,\]
still do not fix the time dependent gauge transformations completely.
There are still `large' periodic gauge transformations wrapping around the
maximal torus $\bT$ which shift $A^{\lt}_{0}$ by elements of the integer
lattice $I$ of $\lt$. Explicitly
(denoting the time parameter by $s$ to avoid confusion)
these gauge transformations can be written as
\bea
&&t(s) = t(0)\exp s\gamma\;\;,\;\;\;\;\;\;t(0)=t(1) \leftrightarrow
\gamma\in I\;\;,\nonumber\\
&&t(s)^{-1}A_{0}\,t(s) + t(s)^{-1}\del_{0}t(s) = A_{0} + \gamma\;\;.
\label{lgt}
\eea
As $\lt/I = \bT$, eliminating these shifts is
tantamount to regarding $A_{0}$ as a {\em compact} scalar field $\f^{\lt}$
taking values in $\bT$. We have thus found that Chern-Simons theory
on $\SS$ is equivalent to an Abelian topological field theory,
\be
Z_{\SS}(S_{CS}) = \int D[\phi,A] \exp(i(k+h)\,S_{\phi F}(\phi,A))\;\;,
\label{phif1}
\ee
with action  and measure given by
\bea
S_{\phi F}(\phi,A)&=&\trac{1}{2\pi}\int_{\S}\tr \phi F_{A}\;\;,
\label{phif2}\\
D[\phi,A] &=& D\phi\,DA\,\det({\bf 1}-{\rm Ad}(e^{i\phi}))^{1-g}
\label{phif3}
\eea
respectively.
Here we have already included the shift $k\ra k+h$, whose occurrence we
will establish in section 6.

\subsection{Comparison with BF Theory}

\noindent
The action (\ref{phif2}) is a `compact' counterpart of the so-called
BF theories, studied extensively in e.g.~\cite{btbf2,ewym,btym,pr,hobf,btbf1},
and defined in any dimension $n$ by
\be
S_{BF} = \trac{1}{2\pi}\int_{M}\tr BF_{A}\;\;,\label{bfact}
\ee
where $B$ is an ordinary (i.e.~non-compact) $(n-2)$-form. It will be
useful to keep in mind the following differences between the compact
and non-compact models in two dimensions:

\begin{enumerate}

\item The non-compact Abelian BF action (or rather $\exp ik S_{BF}$)
also enjoys the invariance $B\ra B + \gamma, \gamma\in I$,
since
\be
\int_{\S}F_{A} \in 2\pi{\bf Z}\;\;. \label{chern}
\ee
The (crucial) difference however is that here this symmetry is
not a consequence of some underlying remnant gauge invariance
but just some global symmetry of the action. As such it need not
and should not be eliminated, and all one can expect is to find
it unitarily represented on the Hilbert space of the theory
(something that is rather trivially true in this case).

\item In two dimensions (and, with some {\em caveat}
also in general, see \cite{btbf1}) the integral over $B$ simply imposes
the delta function constraint  $F_{A}=0$, so that the partition function
calculates the volume of the moduli space of flat connections, with
measure given by the Ray-Singer torsion. In $n=2$ this measure coincides
with the symplectic measure \cite{ewym} and hence, with proper
normalization, the partition function is the symplectic volume of
$\M$. The compactness of $\f$ in (\ref{phif2}) on the other hand
implies that the partition function is no longer a simple delta
function but some deformation thereof. In fact, in terms of a
suitably chosen mode expansion (spectral representation of the
delta function) one finds that sufficiently high modes of
the delta function are cut off due to the compactness of $\f$.

\item Note also that, in the case of BF theories, any prefactor (coupling
constant) like $k$ in the path integral can be absorbed by a rescaling of
$B$, so that the result is essentially independent of $k$.
This is something that, due to the compactness of $\f$, cannot be
done in the action $S_{\f F}$, as a rescaling of $\f$ would change
its radius. We thus expect the partition function (and hence that of \CS\
theory) to depend in a much more subtle manner on $k$, something that
is indeed borne out by the result, the Verlinde formula. One would,
however, expect the large $k$ limit of this result to agree with the
partiton function of BF theory since, by rescaling, the large $k$
limit corresponds to a larger and larger radius of $\f$. This can indeed
be verified and is in
agreement with the expectation that in the semi-classical limit
of \CS\ theory the dimension of the Hilbert space is equal to the
volume (number of cells) of phase space.
\end{enumerate}
%xxx

\section{The ${\bf G/G}$ model from Chern-Simons theory on $\bf\SS$}

The purpose of this and the following sections
is to give an alternative two-dimensional
derivation of the results of the previous section. In particular, this
will allow us to understand the determinant
in (\ref{phif3}) as arising not from the Ray-Singer torsion of $\SS$ but
this time from an infinite dimensional version of the {\em Weyl integral
formula} applied to the partition function of the $G/G$ model. As a preliminary
result we establish the equivalence of Chern-Simons theory on $\SS$
and the $G/G$ model on $\S$ directly at the level of the action and the
path integral. This implies that the partition function of the $G/G$ model
is the dimension of the space of conformal blocks of the $G$ WZW model
and that the fusion rules are reproduced by the correlation functions of
the traces of the group valued fields $g(z,\bar{z})$, something that has been
conjectured and partially verified in e.g.~\cite{verver,sp,ewwzw}.

The strategy will be to write the partition function $Z_{\SS}$ of \CS\
theory as the trace of an amplitude on $\SI$ and to trade $A_{0}$ for the
time independent group valued field $g=P\exp\oint A_{0}dt$,
the holonomy of $A_{0}$, which captures
the entire gauge invariant information carried by $A_{0}$. This is
just the three-dimensional counterpart of the procedure employed in
section 3.1
to determine the torsion of $S^{1}$.
In that way we will end up with a two-dimensional theory which is
expressed solely in terms of $g$ and the (time independent) boundary values of
the spatial components of the connection. In order to recognize this
as the $G/G$ model, we begin by recalling the action of the $G$ WZW model
and the $G/H$ coset models.

\subsection{The Gauged WZW Model}

\noindent In complex coordinates on $\S$, the action of
the WZW model is (we follow the conventions of
\cite{ewwzw})
\bea
S_{G}(g) &=& S_{0}(g)-i\Gamma(g)\;\;,\label{2}\\
S_{0}(g) &=& -\trac{1}{4\pi}\int_{\S}d^{2}\! z \g\dz g \g\dzb g
           \;\;,\label{3}\\
\Gamma(g)&=& \trac{1}{12\pi}\int_{N}d^{3}\!x \epsilon^{ijk}
             \g\del_{i}g \g\del_{j}g \g\del_{k}g\label{4}\;\;,
\eea
where $\del N = \S$, i.e.~$N$ is what is known as a handlebody.
$S_{G}$ is invariant under a $\bG_{L}\times \bG_{R}$
symmetry, $g\ra agb^{-1}$, which (for the particular choice of coefficient
in (\ref{4})), is extended to a $\bG_{L}(\bar{z})\times \bG_{R}(z)$
Kac-Moody symmetry. Only `anomaly-free' subgroups of $\bG_{L}\times \bG_{R}$
can be gauged; these include, in particular, all subgroups ${\bf H}$ of the
adjoint group $\bG_{{\rm adj}}$ ($g\ra aga^{-1}$). Introducing an ${\bf H}$
gauge field $A$, the action $S_{G/H}$ of the gauged WZW model is
\bea
S_{G/H}(g,A) &=& S_{G}(g) + S_{/H}(g,A)\label{5}\;\;,\\
S_{/H}(g,A) &=& -\trac{1}{2\pi}\int_{\S}d^{2}\! z
(\az\dzb g\g -\azb\g\dz g + \az\azb - \g\az g\azb)\;\;.\nonumber
\label{6}
\eea
This action gives a field-theoretic realization
of the GKO coset models \cite{gk}. Taking ${\bf H}=\bG$ in
(\ref{5}) one
obtains the action of the topological $G/G$ model discussed
by Verlinde and Verlinde \cite{verver} and more recently in
\cite{sp,ewwzw}.
For later purposes it will be convenient to have this action written
in terms of differential forms,
\bea
S_{G/G}(g,A)&=&-\trac{1}{8\pi}\int_{\S}\tr \g d_{A}g *\g d_{A}g -i\Gamma(g,A)
\label{diff}\\
\Gamma(g,A)&=& \trac{1}{12\pi}\int_{N}\tr (\g dg)^{3}-\trac{1}{4\pi}\int_{\S}
\tr\left(A(dg\,\g + \g dg) + A\g A g\right)\;\;.\nonumber
\eea

\subsection{Gauge Fixing}

\noindent
To streamline the derivation of $S_{G/G}$ from $S_{CS}$, we make some
preliminary observations. First of all, as we have already seen in
section 2.1, in
discussing \CS\ theory on a three-manifold of the form $\SS$, it is
permissible and convenient to choose the gauge $\del_{0}A_{0}=0$.
In this gauge
\be
g_{t}\equiv P \exp \int_{0}^{t}A_{0} = \exp tA_{0}
\ee
with $g_{0}=1$ and $g_{1}=g$. Note also that $^{g_{t}}A_{0}=0$ where we
have introduced the notation $^{g}A=A^{\g}$.

We will enforce this gauge condition
algebraically as in section 2.1 by imposing the conditions
(\ref{cond}) on the ghost and multiplier fields. The Faddeev-Popov
determinant we obtain in this way will be $\Det'D_{0}$ instead
of the $\Det D_{0}$ of section 2.2, as the conditions (\ref{cond}) are
now imposed on all the Lie algebra components of the ghosts.
As on constant modes $D_{0}$ reduces to ${\rm ad}(A_{0})$,
these two determinants are related by
\be
\Det D_{0} = \Det'D_{0} \times \Det {\rm ad}(A_{0})\;\;.\label{addet}
\ee
The difference between these two determinants will turn out to be
crucial below.

\subsection{Boundary Conditions and Boundary Terms}

\noindent
For the amplitude on $\SI$, with $A_{0}$ regarded as an
external source, we choose the holomorphic representation of the
path integral.
Calling the quantum field $B$, we wish to fix the boundary
conditions
\be
\bz|_{\Sa}=\az\;\;,\;\;\;\;\;\;\bzb|_{\Sb}=\azb\;\;.\label{10}
\ee
In order to do that, we have to add boundary terms to the action,
\be
S_{CS}(A_{0},B)\ra S_{CS}(A_{0},B) -\trac{1}{4\pi}\int_{\Sa}\bz\bzb
                                       -\trac{1}{4\pi}\int_{\Sb}\bz\bzb
\;\;.\label{11}
\ee
The path integral amplitude with these boundary conditions we will denote by
$Z_{\SI}[A_{0};\az,\azb]$. In terms of this, the partition function of
\CS\ theory on $\SS$ can be written as
\be
Z_{\SS} = \int DA_{0}DA\, \Det' D_{0}\, Z_{\SI}[A_{0};\az,\azb]
 \exp(\trac{ik}{2\pi} \int_{\S} \az\azb)\label{13}\;\;,
\ee
where the last factor is the K\"ahler potential measure required
in the holomorphic representation and compensating for the boundary terms
in (\ref{11}).

Finally, we need to keep track of the fact that on a manifold with boundary
$\exp ik S_{CS}$ is not invariant under arbitrary gauge transformations.
Rather, one has
\be
\exp ik S_{CS}(\unA^{h}) = \Theta(\unA,h)^{k} \exp ik S_{CS}(\unA)
\;\;,\label{8}
\ee
where
\be
\Theta(\unA,h) = \exp(- \trac{i}{12 \pi} \int_{M} (h^{-1}dh)^{3} +
\trac{i}{4\pi} \int_{\partial M}  \unA dh h^{-1}) \label{9}
\ee

As a consequence, the amplitude transforms non-trivially under gauge
transformations and for $\SI$ and $g_{t}=\exp tA_{0}$ one finds
\be
Z_{\SI}[A_{0};\az,\azb^{g}]=Z_{\SI}[^{g_{t}}A_{0}=0;\az,\azb]\,C[\azb,g]\;\;,
\label{18}
\ee
where the Polyakov-Wiegmann cocycle $C[\azb,g]$ is
\be
C[\azb,g] = \exp ik S_{G/G}(\g,\az=0,\azb)\;\;.
%C[\azb,g]=\exp\left(-\trac{ik}{12\pi}\int_{M}(\g dg)^{3}-\trac{ik}{4\pi}
%\int_{\Sb}(\g\dz g\g\dzb g +2\azb\dz g\g)\right)
\label{19}
\ee
This is nothing but the familiar transformation behaviour of Chern-Simons
wave functionals under gauge transformations, see e.g. \cite{emss}.

\subsection{The Measure}

\noindent
We would also like to convert the linear measure $DA_{0}$ on the
Lie algebra to the Haar measure $Dg$ on the group. It follows
from Duhamel's formula
\be
\exp(-X)\,\d\exp(X) = \int_{0}^{1}\!ds\,\exp(-sX)\,\d X\,\exp(sX)\;\;,
\ee
that these are related non-trivially by \cite{bgv}
\be
Dg = \Det\left(\frac{({\bf 1}-{\rm Ad}(e^{A_{0}}))}
{{\rm ad}(A_{0})}\right)\, DA_{0} \label{haar}\;\;.
\ee
It is quite remarkable that, as a consequence of (\ref{addet}) and
the calculations of section 3, $\Det'D_{0}$ provides precisely
this conversion factor (the missing $\Det'\del_{0}$ being kindly
supplied by the $B$ integration below).

\subsection{Synthesis}

\noindent
Putting all this together and using the gauge invariance of the gauge
field measure one finds that
\be
Z_{\SS}=\int DgDA Z_{\SI}[A_{0}=0;\az,\azb]
C[\azb,g]\exp(\trac{ik}{2\pi}\int_{\S}\az\azb^{g})\;\;.
\label{14}
\ee

Thus, finally, the remaining path integral we need is
$Z_{\SI}[A_{0}=0;\az,\azb]$ which is not a functional of $g$ and may
be straightforwardly evaluated, providing due care of the boundary data
is taken. One can introduce these boundary conditions into the
path integral via Lagrange multipliers and then solve the resulting
equations of motion for $B$. Alternatively,
one uses the fact, that the boundary
terms in $Z_{\SI}[A_{0}=0;\az,\azb]$ were designed to cancel those
arising from the variation of the \CS\ action, so that one can read off
directly the equations of motion $\del_{0}B=0$ which (with the
boundary conditions (\ref{10})) imply $B=A$. Either way one
obtains
\be
Z_{\SI}[A_{0}=0;\az,\azb]=N \exp(-\trac{ik}{2\pi}\int_{\Sa}\az\azb)\;\;.
\label{20}
\ee
Here $N^{-1}$
is the factor $\Det'\del_{0}$ which we already absorbed
into the proper normalization of the measure (\ref{haar}).
Putting all the pieces together from equations
(\ref{18}-\ref{20}) one obtains from the \CS\ path integral
on $\SS$ precisely  the $G_{k}/G_{k}$ gauged WZW action on $\S$ (\ref{5})
(with $g\ra\g$, which is due to our choice of orientation for $\SI$),
\bea
S_{G/G}(\g,A) &=& \underbrace{S_{G}(\g) -\trac{1}{2\pi}\int_{\S}d^{2}\! z
(\azb\dz g\g}_{(\ref{19})} + \underbrace{\az\azb}_{(\ref{20})} -
\underbrace{\az\azb^{g}}_{(\ref{14})})\;\;,
\eea
with the correct Haar measure $Dg$.
It is also evident from this derivation, that correlators of `vertical
Wilson loops' in Chern-Simons theory (corresponding to the fusion
rules) at level $k$ are equal to the correlators of
${\rm tr}\;g(z,\bar{z})$ in the $G_{k}/G_{k}$ coset model.

\section{Abelian Reduction of the ${\bf G/G}$ Theory}

The next task is to evaluate the partition function of the $G/G$ model
obtained in the previous section. We will do this by making use of the
Weyl integral formula. This formula, whose precise form and derivation
we will recall below, relates the integral of a conjugation
invariant function on a copmact
group $\bG$ to an integral over the maximal torus
$\bT$,
an Abelian group. Applied to the path integral of the $G/G$ model it thus
permits one to effectively reduce the path integral to that of an Abelian
theory which can be exactly calculated.

Before proceeding we want to point out that the
range of applicability of the
Weyl integral formula and its
relatives (valid e.g.~for Lie algebras or non-compact semi-simple Lie
groups) to path integrals is not limited to the $G/G$ models
considered here but can also be used to significantly simplify
the evaluation of path integrals in other theories with local symmetries.
We just mention that this provides possibly the shortest
available derivation of the partition function of Yang-Mills theory
on an arbitrary closed surface \cite{btfp},
obtained previously in e.g.~\cite{ewym,btym}
by various other methods.

\subsection{The Weyl Integral Formula}

To write down and explain the \wif\ we will have to introduce some notation.
Thus let $\bG$ be a compact Lie group (which we will also assume to be
semi-simple and simply connected later on) and $\bT$ a maximal torus of $\bG$.
The rank $r={\rm rk}(\bG)$ of $\bG$ is the dimension of $\bT$.
Any element of $\bG$ can be conjugated into $\bT$ (`diagonalized')
and the residual conjugation
action of $\bG$ on $\bT$ (permutation of the diagonal entries)
is that of a finite group, the Weyl group
$W=N(\bT)/\bT$ ($N(\bT)$ denotes the normalizer of $\bT$ in $\bG$). Furthermore
any two maximal
tori are conjugate to each other and the set $\bG_{r}$ of regular
elements of $\bG$ (i.e.~those whose centralizer is conjugate to $\bT$)
is open and dense in $\bG$. It follows that the conjugation map
\bea
\bG/\bT \times \bT_{r} &\ra& \bG_{r} \nonumber\\
(g,t) &\mapsto& \g t g \label{w1}
\eea
is a $|W|$-fold covering onto $\bG_{r}$. Corresponding to a choice of $\bT$
we have a direct sum decomposition of the Lie algebra
$\lg$ of $\bG$, $\lg = \lt \oplus \lk$, orthogonal with respect to the
Killing-Cartan metric on $\lg$. $\bG$ acts on $\lg$ via the adjoint
representation $\rm Ad$. This induces an action of $\bT$ which acts trivially
on $\lt$ and leaves $\lk$ invariant (the isotropy representation
${\rm Ad}_{\lk}$ of $\bT$ on $\lk$). Thus the compexified Lie algebra
$\lg_{\bf C}$ splits
into $\lt_{\bf C}$ and the one-dimensional eigenspaces $\lg_{\a}$ of the
isotropy representation, labeled by the roots $\a$
($\overline{\lg_{\a}}=\lg_{-\a}$) and one obtains the Cartan
decomposition
\be
\lg_{\bf C} = \lt_{\bf C} \oplus \sum_{\a} \lg_{\a}\;\;.\label{w6}
\ee

On $\bG$ and $\bT$ there
exist natural invariant Haar measures $dg$ and $dt$
normalized to $\int_{\bG}dg= \int_{\bT}dt = 1$. For the purpose of
integration over $\bG$ we may restrict ourselves to $\bG_{r}$ and we can
thus use (\ref{w1}) to pull back the measure $dg$ to $\bG/\bT \times \bT$.
Calculating the corresponding Jacobian one finds the Weyl integral
formula
\be
\int_{\bG}\!dg\;f(g) =\trac{1}{|W|}\int_{\bT}\!dt\; \det\dw(t)
                    \int_{\bG/\bT}\!dg\;f(\g t g)\;\;,\label{w2}
\ee
where
\be
\dw(t) =  ({\bf 1}_{\lk}- {\rm Ad}_{\lk}(t)) \label{w3}\;\;.
\ee
In particular, if $f$ is conjugation invariant (a class function), it is
determined entirely by its restriction to $\bT$ (where it is $W$-invariant)
and (\ref{w2}) reduces to
\be
\int_{G}\!dg\;f(g) = \trac{1}{|W|}
           \int_{\bT}\!dt\; \det\dw(t) f(t) \;\;, \label{w4}
\ee
which is the version of the \wif\ which we will make use of later on.
Note that the determinant of $\dw(t)$ is precisely equal to the
Ray-Singer torsion on the circle we derived in section 3, where we
also sketched a physical explanation for this coincidence.

It follows from (\ref{w6}) that
\be
\det({\bf 1}-{\rm Ad}_{\lk}(t)) = \prod_{\a}(1-e^{\a}(t))\;\;.\label{w7}
\ee
Decomposing the set of roots into positive ($\a > 0$) and negative roots
and introducing the Weyl vector $\rho = \frac{1}{2}\sum_{\a > 0} \a$,
this can also be written in terms of the denominator $Q(t)$ of the
Weyl character formula,
\be
Q(t)= \sum_{w\in W}\det(w)e^{w(\rho)}(t)\;\;,\label{w8}
\ee
as
\be
\det({\bf 1}-{\rm Ad}_{\lk}(t)) = Q(t)\overline{Q(t)} \;\;.\label{w9}
\ee

\subsection{Example: $SU(2)$}

\noindent
Let us illustrate the above in the case when $\bG=SU(2)$.
We parametrize elements
of $SU(2)$ and $\bT=U(1)$ as
\be
g = \mat{g_{11}}{g_{12}}{g_{21}}{g_{22}}\;\;,\;\;\;\;\;\;
t = \mat{e^{i\vf}}{0}{0}{e^{-i\vf}}\;\;.                     \label{w10}
\ee
The Weyl group $W={\bf Z}_{2}$ acts on $\bT$ as $t\mapsto t^{-1}$.
We use the trace to identify the Lie algebra $\lt$ of $\bT$ with its dual
and introduce the positive root $\a$ and the fundamental weight $\la$,
\be
\a = \mat{1}{0}{0}{-1}\;\;,\;\;\;\;\;\;\la =\trac{1}{2}\mat{1}{0}{0}{-1}\;\;,
\ee
satisfying the relations
\be
\tr \a^{2} = 2 \;\;,\;\;\;\;\;\;\tr \a\la = 1\;\;,\;\;\;\;\;\;
\rho = \trac{1}{2}\a = \la\;\;.
\ee
Later on we will find it convenient to parametrize elements of $\bT$ in terms
of weights. Thus, we write
$t= \exp i\la\phi$, where $\f$ is related to $\vf$ by $\f=2\vf$.
Then the expression $\exp(\a)(t)$
entering (\ref{w7}) becomes $\exp(\a)(t) = \exp i\f$, and
the Weyl denominator (\ref{w8}) and the determinant (\ref{w3}) are
\bea
Q(t)&=& 2i\sin\f/2\;\;,\nonumber\\
\det\dw(t) &=& 4 \sin^{2}\f/2\;\;.                            \label{w11}
\eea
Hence the \wif\ for class functions is  (with $f(\f)\equiv f(\exp i \la\f)$)
\bea
\int_{\bG}\!dg\;f(g) &=& \frac{1}{2}\int_{0}^{4\pi}\!\frac{d\f}{4\pi}4
\sin^{2}(\f/2) f(\f)\nonumber\\
&=& \frac{1}{2\pi}\int_{0}^{4\pi}\!d\f\; \sin^{2}(\f/2) f(\f)\nonumber\\
&=& \frac{1}{\pi}\int_{0}^{2\pi}\!d\f\; \sin^{2}(\f/2) f(\f)\;\;. \label{w12}
\eea
Here the last line follows e.g.~from writing $2\sin^{2}(\f/2)=
1-\cos \f$ and
is a useful reformulation because it effectively incorporates the action
of the Weyl group.

\subsection{Faddeev-Popov Derivation}

\noindent
We mention in passing that
these formulae can be obtained \`a la Faddeev-Popov by `gauge fixing' the
non-torus part of $g$ to zero (i.e.~by imposing $g\in\bT$ as a gauge
condition). This amounts to inserting $1$ in the form
\be
1=\trac{1}{|W|}\int_{\bG/\bT}\!dh\;\int_{\bT}\!dt\;\d(h^{-1}gh t^{-1})
\det\dw(t)
\label{w5}
\ee
into the integral on the lhs of (\ref{w2}) and performing the integral
over $g$,
\bea
\int_{\bG}\!dg\;f(g) &=&
\trac{1}{|W|}\int_{\bG}\!dg\;\int_{\bG/\bT}\!dh\;\int_{\bT}\!dt\;
\d(h^{-1}gh t^{-1})\det\dw(t)\nonumber\\
&=&\trac{1}{|W|}\int_{\bT}\!dt\;\det\dw(t)\int_{\bG/\bT}\!dh\;f(h^{-1}th)
\;\;.
\eea
The Faddeev-Popov determinant $\det\dw(t)/|W|$ can then be
obtained directly from the (BRST) variation of the condition $g\in\bT$.
In the $SU(2)$ case this amounts to fixing the gauge
$g_{12}=g_{21}=0$. Since
infinitesimally $g_{12}$ transforms under conjugation as ($a\in\lg$)
\be
\d g = [g,a] \Rightarrow \d g_{12} = 2i a_{12} \sin\vf\label{w13}\;\;,
\ee
the resulting Faddeev-Popov determinant is just (\ref{w11}), while the
additional factor of $1/2$ accounts for the residual gauge freedom
(conjugations leaving $\bT$ invariant).

\subsection{Abelianization of the WZW model}

It is now straightforward to apply the Weyl integral formula to the
partition function of the $G/G$ model. As the functional of $g$ that
one obtains after having performed the path integral over the gauge
fields is locally and pointwise conjugation invariant,
\be
{\cal F}(g)\equiv\int DA \exp(ik S_{G/G}(g,A)) = {\cal F}(h^{-1}gh)
\label{g1}\;\;,
\ee
one can formally use the Weyl integral formula pointwise to reduce the
remaining path integral over the group valued fields to one over fields
taking values in the torus $\bT$ of $\bG$. Or, in other words, one can use
the gauge invariance of the $G/G$ action to impose the gauge condition
$g\in \bT$. Either way one will generate a functional version of the
determinant $\det\dw(t)$ we encountered
in section 5.1,
\be
\int Dg {\cal F}(g)=
\int Dt DA\,{\rm Det}\,({\bf 1}-{\rm Ad}_{\lk}(t))
\exp(ik S_{G/G}(t,A))\;\;.\label{g2}
\ee

Let us now in turn take a look at the two parts $S_{G}(t)$ and $S_{/G}(t,A)$
of the action $S_{G/G}(t,A)$ defined in (\ref{5}).
For notational simplicity only we will assume in the following that
$\bG=SU(n)$. It should be apparent how to extend this to arbitrary
compact $\bG$.
We will also simplify things as in
section 5.2 by identifying $\lt$ with its dual $\lt^*$. In
particular, we will regard the roots of $\bG$ as elements of $\lt$ and a set
$\{\a_{l}, l = 1,\ldots,n-1 = r\}$ of simple roots as a basis
of $\lt$. We thus expand the gauge field as $A^{\lt} = i\a_{l}A^{l}$. It
is then convenient to parametrize the torus valued field $t$ in terms
of the dual basis $\{\la^{l}\}$ of {\em fundamental weights},
\bea
A^{\lt}&=&i\a_{l}A^{l}\nonumber\\
t&=&\exp i\f^{\lt}\;\;,\;\;\;\;\;\;\f^{\lt} = \f_{l}\la^{l}\;\;.\label{para}
\eea
It is clear from this description that the $\f_{l}$ are compact
scalar fields. We will determine their range (radii) in section 7.4
below when we take into account the effect of the Weyl group.

In the WZW action $S_{G}(t)$ the
kinetic term $S_{0}(t)$ reduces to the standard
kinetic term
\be
S_{0}(t) = \trac{1}{4\pi}\int_{\S}  \la^{kl}
            \dz\f_{k}\dzb\f_{l}\label{g3}
\ee
for compact bosons. Here $\la^{kl}={\rm tr}(\la^{k}\la^{l})$ is
the inverse of the Cartan matrix.

The WZ term $\Gamma(t)$ does not vanish,
as one might naively expect in the case of Abelian groups, but
as a `topological' term it only depends on the winding numbers
of the field $\f$. The reason for the appearance of this contribution
is, that maps from $\S$ to $\bT$ with non-trivial windings cannot
necessarily be extended to the interior $N$ of $\S$ {\em within} $\bT$,
as some (half) of the non-contractable cycles of $\S$ become contractible
in the handlebody $N$. The general form of this term is \cite{gacs}
\be
\Gamma(t) = \int_{\S} \mu^{kl}\,d\f_{k}\,d\f_{l} \;\;,
\ee
where $\mu^{kl}$ is some antisymmetric matrix. As we will show below
(cf.~the discussion after (\ref{bf2act}))
that the non-trivial winding sectors do not contribute to the partition
function, we will not have to be more precise about this term here.

\subsection{The Gauge Field Contribution}

\noindent
In the action $S_{/G}(t,A)$ the contributions from the $\lt$ and $\lk$
components $A^{\lt}$ and $A^{\lk}$ of the gauge field $A$ are neatly
seperated so that it is easy to perform the path integral over the $A^{\lk}$,
leaving behind an effective Abelian theory. In fact,
because $\lt$ and $\lk$ are orthogonal to each other with respect to the
invariant scalar product (trace), only $A^{\lt}$ will contribute to
the terms of the form $\azb t^{-1}\dz t$ and $\az \dzb t \,t^{-1}$
(cf.~equation (\ref{5})),
\be
-\trac{1}{2\pi}\int_{\S} (\az\dzb t \,t^{-1} - \azb t^{-1}\dz t) =
\trac{1}{2\pi}\int_{\S}( \az^{l}\dzb\f_{l} - \azb^{l}\dz\f_{l})
\;\;.\label{g5}
\ee
We now observe that we can eliminate (\ref{g3}) altogether by
a shift of the gauge field,
\be
A^{\lt}\ra A^{\lt} + \trac{1}{4}*d\f^{\lt}\;\;, \label{g6}
\ee
(note that this is not a gauge transformation)
leaving us with the simple action
\be
\trac{1}{2\pi}\int_{\S}\tr A\,d\f\;\;.\label{act}
\ee
As we will see in section 7.2 that only the constant modes of $\f$ contribute
to the path integral, we could have just as well carried the term
(\ref{g3}) along until the end. And while this would have avoided the seeming
nuisance of a metric dependent field redefinition, it is nicer to
work with the action (\ref{act}) because of its resemblance to
other topological gauge theories in two dimensions.

We would now like to integrate by parts in (\ref{act}) to put it
into the form of the action of a BF theory (\ref{bfact}),
whose action in $2d$, we recall, is
\be
S_{BF}=\trac{1}{2\pi}\int_{\S}\tr BF_{A}\;\;,\label{bf2act}
\ee
where $B$ is an ordinary (non-compact) scalar field.

At first, the
compactness of $\f$ may cast some doubt on this procedure since,
with $\f$ being an `angular variable', $d\f$ is not necessarily exact. One
would therefore expect to pick up `boundary' terms from the monodromy of $\f$.
The following argument shows that in the $G/G$ model (and hence in
Chern-Simons theory on $\SS$) the non-trivial winding sectors of
these fields do not contribute to the partition function: as it is only
the harmonic modes of $A$ that couple to the non-exact (winding) parts
of $d\f$, integration over these modes will set the non-zero
winding modes of $\f$ to zero. As there is no Jacobian involved in
going from $A$ to `harmonic modes plus rest', this shows that we can
indeed integrate by parts in (\ref{act}) (with the understanding that
the harmonic modes of $A$ no longer appear) and we thus arrive at the
BF like action
\be
S_{\f F} = \trac{1}{2\pi} \int_{\S}\f_{l}F^{l}\;\;.\label{g7}
\ee
Here $F^{l}=dA^{l}$ is the curvature of the Abelian gauge field $A^{l}$.
This argument also takes care of the WZ term and the result
is in pleasant agreement with that  obtained
by quite different means in (\ref{phif2}).
One of the important differences between this theory and
the ordinary BF models
is of course, as already pointed out in section 3.6, that here
the scalar fields $\f_{l}$ are compact which implies that the
integral over them will not simply produce a delta function onto flat
connections as is the case in the non-compact BF theories.

While $A^{\lk}$ has made no appearance in the above, it is $A^{\lt}$
that will drop out of the remaining term
$\az\azb - \g\az g\azb$ which becomes simply
\be
\az\azb - \g\az g\azb \ra \az^{\lk}({\bf 1}-{\rm Ad}_{\lk}(t))\azb^{\lk}
\;\;.\label{g4}
\ee
Thus the $A^{\lk}$ integral will give rise to a determinant that {\em
formally} (cf.~the considerations in section 2.2)
cancels against the Faddeev-Popov determinant in (\ref{g2}).
Of course this is not correct, as certainly the zero modes will leave
behind a finite dimensional determinant. Furthermore, the determinants
should be properly regularized, and we will perform this in the
following section. Suffice it to say here that this gives rise to
the shift $k\ra k+h$. In fact, the residual finite dimensional determinant
$\det^{1-g}\dw(t)$ and the shift will arise simultaneously as the
gravitational and gauge field contributions to the chiral anomaly.

Anticipating this result we have thus deduced that the $G/G$ model on
$\S$ (and hence Chern-Simons theory on $\SS$) is equivalent to the
two-dimensional Abelian $\f F$ theory (\ref{g7}) with measure
$\det^{1-g}\dw$. In particular this is in perfect agreement with
the results of section 3 (cf.~equations (\ref{phif1}-\ref{phif3})).

\section{The shift ${\bf k\ra k+h}$}

Thus far we have cancelled the determinants arising from the gauge field
integration against those of the ghosts, up to harmonic modes, with gay
abandon. This was the case e.g.~in sections 2 and 3, where we claimed that
the ratio (\ref{ratio}) of determinants would give rise to
the Ray-Singer torsion of $\SS$ and where we also promised that a gauge
invariant regularization would produce the shift $k\ra k+h$. This was
also the case in the previous section, where we argued that the ratio
of the determinants $\Det\dw(t)$ leads to the same result.
It is thus high time to give a precise meaning to the ratios of
determinants involved and to declare the regularisation that is
used to do this. We start by considering the determinants that
arise in the $G/G$ model. As we have shown the equivalence of \CS\ theory
with the $G/G$ model, this also takes care of the former. However, a
small additional argument allows
us to reduce the calculation of the \CS\ determinants directly to the $G/G$
result and the calculation of the Ray-Singer torsion on $S^{1}$ and we
indicate in section 6.4 how this is done.

\subsection{The Dolbeault Complex in the $G/G$ Model}

The first thing to note is that our gauge fixing so far has only been
partial as we have been careful to preserve the Abelian $\bT$ invariance.
We should thus regularize in a manner which respects this residual
gauge invariance and we will accomplish this by using a heat kernel
(or $\zeta$-function) regularization based on the $\lt$ covariant
Laplacian $\Delta_{A} = -(d_{A}^{*}d_{A}+d_{A}d_{A}^{*})$ where
$A$ is the $\bT$ gauge field. For an operator $\cal O$ we set
\be
\log\Det{\cal O} = \Tr e^{-\eps\Delta_{A}}\log{\cal O}\;\;,
\ee
where we use $\Tr$ to denote a functional trace (e.g.~including
an integration).

We begin with the determinant that arises on integrating out
$A^{\lk}$, see (\ref{g4}). Up to an overall factor,
the relevant part of the action is
(using the differential form version (\ref{diff}))
\be
\tr( A^{\lk}*A^{\lk} - A^{\lk}t^{-1}(i+*)A^{\lk}\,t)\;\;.\label{trk}
\ee
To put this into a more explicit form we recall that, on the
root space $\lg_{\a}\ss\lk_{\bf C}$, ${\rm Ad}(t)$ acts by multiplication
by $\exp i\a(\f)$. Furthermore, with respect to the Killing-Cartan
metric (trace), $\lg_{\a}$ and $\lg_{\beta}$ are orthogonal unless
$\beta = -\alpha$. Thus, expanding $A^{\lk}$ in terms of basis vectors $E_{\a}$
of $\lg_{\a}$ such that
\be
A^{\lk}=\sum_{\a}E_{\a}A^{\a}\;\;,\;\;\;\;\;\;\tr(E_{\a}E_{-\a}) = 1\;\;,
\ee
we can break up (\ref{trk}) into a
sum of terms depending only on the pair $\pm\a$. We obtain
\bea
(\ref{trk}) &=& \sum_{\a}A^{\a}*A^{-\a}-A^{\a}e^{-i\a(\f)}(i+*)A^{-\a}
\nonumber\\
&=& \sum_{\a>0}\left[ A^{\a}(i+*)M_{-\a}A^{-\a}
           - A^{\a}(i-*)M_{\a}A^{-\a}\right]\;\;,
\eea
where $M_{\a}$ is the number
\bea
M_{\a}&=&\left(1-e^{i\a(\f)}\right)\;\;,\nonumber\\
\prod_{\a}M_{\a}&=&\det ({\bf 1}-{\rm Ad}_{\lk}(e^{i\f}))\;\;.\label{s9}
\eea
Writing this in terms of the scalar product (\ref{scal}) on $1$-forms,
one sees that the path integral over $A^{\lk}$ yields
\be
\prod_{\a>0}\Det\left[(1+i*)M_{\a} + (1-i*)M_{-\a}\right]^{-1}\;\;.
\label{s1}
\ee
Here we recognize the projectors
\be
P_{\pm} = \trac{1}{2}(1\pm i*)
\ee
onto the spaces of $(1,0)$-forms ($\sim dz$) and $(0,1)$-forms
($\sim d\bar{z}$) respectively and thus (\ref{s1}) exhibits
quite clearly the chiral nature of the (gauged) WZW model. As a
consequence of the presence of the projectors $P_{\pm}$ in (\ref{s1}),
the two summands act on different spaces. For each $\a$ we may thus
write the determinant as a product of the $(1,0)$ and $(0,1)$
pieces,
\be
\Det\left[(1+i*)M_{\a} + (1-i*)M_{-\a}\right]^{-1}
=\left[\Det_{(1,0)}M_{\a}\right]^{-1} \times
\left[\Det_{(0,1)}M_{-\a}\right]^{-1}\;\;.\label{s2}
\ee

Before evaluating this, we will combine it with the contributions
from the ghosts (equivalently, the Weyl integral formula). The ghost
action has the form
\be
\sum_{\a>0}\left[\bar{c}^{\a}*M_{-\a}c^{-\a}
    + \bar{c}^{-\a}*M_{\a}c^{\a}\right]     \label{s3}\;\;,
\ee
and therefore the ghost determinant is
\be
\prod_{\a>0}\Det_{0}M_{\a}\Det_{0}M_{-\a}\label{s4}\;\;.
\ee
Combining this with (\ref{s2}), we see that we need to determine
and make sense of
\be
\prod_{\a>0}\left[\Det_{0}M_{\a}\Det^{-1}_{(1,0)}M_{\a}\right]
          \left[\Det_{0}M_{-\a}\Det^{-1}_{(0,1)}M_{-\a}\right]\;\;.
\label{s5}
\ee
This we will accomplish by relating the products of these determinants
to the Witten index of the Dolbeault complex. Indeed, suppose that $M_{\a}$
is a constant. Then
\be
\log\Det_{0} M_{\a}\Det^{-1}_{(1,0)}M_{\a} =
\left[\Tr_{0}e^{-\eps\Delta_{A}} - \Tr_{(1,0)}e^{-\eps\Delta_{A}}\right]
\log M_{\a}\;\;,\label{s6}
\ee
where we need to remember that the Laplacian $\Delta_{A}$ acts to the
right on one-forms taking values in $\lg_{(-\a)}$,
the root space of $(-\a)$. There we have
\be
d_{A}|_{(-\a)} = d -i\a (A) \equiv d + \tr(\a\a_{l})A^{l}\;\;,
\ee
so that the `charge' is $\tr(\a\a_{l})$.
The term in brackets is nothing but the index of the Dolbeault complex,
\be
\left[\Tr_{0}e^{-\eps\Delta_{A}} - \Tr_{(1,0)}e^{-\eps\Delta_{A}}\right]
=\sum_{p=0}^{1}(-1)^{p}b^{p,0} = {\rm Index}\;\bar{\del}_{A}\;\;.
\ee
This index can of course be calculated directly from the heat kernel
expansion, but if one does not want to reinvent the wheel one may
call upon the known result that for the Dolbeault operator coupled to
a vector bundle $V$ with connection $A$ one has (see e.g.~\cite{gi})
\be
{\rm Index}\;\bar{\del}_{A} = \int_{M}{\rm Td}(T^{(1,0)}(M)){\rm ch}(V)\;\;.
\ee
In two dimensions this reduces to
\be
{\rm Index}\;\bar{\del}_{A}=\trac{1}{2}\c(\S) + c_{1}(V)\;\;. \label{dolind}
\ee
Therefore, in the case at hand, one finds that (\ref{s6}) equals
\bea
{\rm Index}\;\bar{\del}|_{(-\a)} \log M_{\a} &=& \left[\trac{1}{2}\c(\S) +
c_{1}(V_{(-\a)})\right]\log M_{\a}\nonumber\\
&=&\left[\trac{1}{8\pi}\int_{\S}R + \trac{1}{2\pi}\int_{\S}\tr(\a\a_{l})F^{l}
\right]\log M_{\a}\;\;.
\eea

When $M_{\a}$ is not a constant, one simply has to move $\log M_{\a}$
into the integral, so that one obtains
\be
\log\Det_{0} M_{\a}\Det^{-1}_{(1,0)}M_{\a} =
\trac{1}{8\pi}\int_{\S}R\log M_{\a} +
 \trac{1}{2\pi}\int_{\S}\tr(\a\a_{l})F^{l}\log M_{\a}\;\;.\label{s7}
\ee
To see that this is correct, we write
\bea
\Tr\log M_{\a} e^{-\eps\Delta_{A}} &\equiv& \int\!dx\,
\langle x|\log M_{\a} e^{-\eps\Delta_{A}}|x\rangle\nonumber\\
&=&\int\!dx\,\log M_{\a}(x)\langle x|e^{-\eps\Delta_{A}}|x\rangle\;\;,
\eea
and note that $R$ and $F$ arise as the first Seeley coefficients in the
expansion of $\langle x|e^{-\eps\Delta_{A}}|x\rangle$. It is worthwhile
remarking that
the result (\ref{s7}) is {\em finite}, the $1/\eps$ poles cancelling
between the scalar and one-form contributions. This is
another manifestation of the supersymmetry discussed in section 2.
Looking at the relevant gauge field and ghost terms in the action,
(\ref{trk}) and (\ref{s3}), we see that in the $G/G$ model it
appears in the (chiral) form
\be
\d A^{-\a} = \eps c^{-\a}\;\;,\;\;\;\;\;\;\d\bar{c}^{\a} = *(\eps
(-i+*)A^{\a})
\ee
(with a similar expression for the other components). Again, the
presence of this supersymmetry implies (formally) that only the
finite dimensional spaces of zero modes contribute to the ratio of
determinants.

One can proceed analogously for the second factor in (\ref{s5}). In this
case it is the index of $\del_{A}$ that makes an appearance and which
differs by the sign of the second summand from (\ref{dolind}),
\be
{\rm Index}\;\del_{A} = \trac{1}{2}\c(\S) - c_{1}(V)\;\;.
\ee
As the Laplacian still acts on $\lg_{(-\a)}$, one obtains a
$\log M_{\a} - \log M_{-\a}$ contribution to the gauge field part of the
index, while it is the sum of the two terms that contributes to
the gravitational part. Hence one finds that the regularized determinant
(\ref{s5}) is
\be
(\ref{s5}) = \prod_{\a>0}
\exp\left(\trac{1}{8\pi}\int_{\S}R\log M_{a}M_{-\a}
+ \trac{1}{2\pi}\int_{\S}\tr(\a\a_{l})F^{l}\log\frac{M_{\a}}{M_{-\a}}
\right)\;\;.
\label{s8}
\ee
We will now consider seperately the two contributions to this expression.

\subsection{Generalized Ray-Singer Torsion}

\noindent
Rewriting the term in (\ref{s8}) that depends on the curvature as
\be
\exp\left[\trac{1}{8\pi}\int_{\S}R\sum_{\a>0} \log M_{\a} M_{-\a}\right]\;\;,
\label{s10}
\ee
one recognizes it as a dilaton like coupling to the metric, the role of the
dilaton being played by
\be
\Phi = \sum_{\a>0}\log M_{\a} M_{-\a} = \log\det({\bf 1} - {\rm Ad}(e^{i\f}))
\;\;.
\ee
For our purposes, however, the most useful way of looking at (\ref{s10}) is
to regard it as a (metric dependent) generalization of the
(metric independent) Ray-Singer torsion
on $\SS$ to non-flat connections. Indeed, $A_{0}dt$ is flat iff $A_{0}$ is
constant iff $\f$ is constant iff $\Phi$ is constant, and in that case
(\ref{s10}) reduces to
\bea
\exp\c(\S)\Phi/2 &=& \left[\det({\bf 1} - {\rm Ad}(e^{i\f}))\right]^{\c(\S)/2}
\nonumber\\&=& T_{\SS}(A_{0})\;\;.
\eea
On the other hand, when $A_{0}$ is not flat on $\SS$, we can think of each
point $x\in\S$ as indexing a flat connection on $S^{1}$. From this point
of view, (\ref{s10}) is an averaging over the different $S^{1}$ connections
weighted by the curvature. This is, of course, not metric independent in
general but turns out to be so for flat connections on $\SS$.

We will show in section 7.2 that only constant $\f$
configurations contribute to the path integral. Hence the expression
(\ref{s10}) will then indeed collapse to the Ray-Singer torsion, as
anticipated by our more naive considerations in sections 3 and 5.

\subsection{The Shift $k\ra k+h$}

\noindent
We now come to the crux of the matter. The second term in (\ref{s8})
is responsible for the shift in $k$. To see this note that
\be
\frac{M_{\a}}{M_{-\a}} = \frac{1-e^{i\a\f}}{1-e^{-i\a\f}} = - e^{i\a\f}
\;\;,
\ee
so that
\be
\prod_{\a>0}\exp\left(\trac{1}{2\pi}\int_{\S}\tr(\a\a_{l})F^{l}\log
  \frac{M_{\a}}{M_{-\a}}\right) = \exp\left[\trac{i}{2\pi}\sum_{\a>0}
\int_{\S}\tr(\a\a_{l})\a(\f)F^{l}\right]\;\;. \label{s11}
\ee
Here we have suppressed the imaginary contribution to the $\log$, as it
will make no appearance for simply connected groups (where $\rho =
\frac{1}{2}\sum_{\a >0}\a$ is integral).

We now put the exponent in more manageable form by noting that the
(negative of the) Killing-Cartan metric $b$ of $\lg$, restricted to $\lt$,
\be
b(X,Y) = - \tr {\rm ad}(X)\,{\rm ad}(Y)\;\;,
\ee
can be written in terms of the roots as
\be
b(X,Y) = 2\sum_{\a>0}\a(X)\a(Y)\;\;.
\ee
Moreover, with our convention that ${\rm ad}(X)|_{\a} = i\a(X)$,
$b(X,Y)$ is related to the Coxeter number (or quadratic Casimir of the
adjoint representation) $h$   via
\be
b(X,Y)= 2h \tr(XY)
\ee
($h=n$ for $SU(n)$). Hence the exponent becomes
\bea
\trac{i}{2\pi}\sum_{\a>0}\int_{\S}\tr(\a\a_{l})\a(\f)F^{l} &=&
\trac{i}{4\pi}\int_{\S} b(\f,\a_{l})F^{l}\nonumber\\
&=& \trac{ih}{2\pi}\int_{\S}\f_{l}F^{l}\;\;,
\eea
which produces precisely the long awaited
shift $k\ra k+h$ in the action $S_{\f F}$,
\be
\trac{ik}{2\pi}\int_{\S}\f_{l}F^{l} \longrightarrow
\trac{i(k+h)}{2\pi}\int_{\S}\f_{l}F^{l}\;\;.\label{shift}
\ee

\subsection{The Calculation for Chern-Simons Theory}

\noindent
We now indicate briefly how the calculation of the determinants
arising in Chern-Simons theory on $\SS$ can be reduced to those performed
for the torsion of the circle $S^{1}$ in section 3.1 and for the $G/G$ model
above. The ratio of determinants that needs to be regularized is given in
(\ref{ratio}). To make our calculation as easy as possible and to make
contact with the $\zeta$-function regularization of section 3 and the
heat-kernel regularization of of the $G/G$ model, we use
a hybrid regularization.

As only differential forms of the type $\Omega^{*}(\S)\ot\Omega^{0}(S^{1})$
and time derivatives enter into (\ref{ratio}), it is convenient to
expand the forms as Fourier series. Then the higher modes can be dealt
with essentially as in the case of the torsion of the circle so that one
finds e.g.
\be
\frac{\Det_{\lt}'\del_{0}|_{\Omega^{0}(\S)\ot\Omega^{0}(S^{1})}}
     {\Det_{\lt}'^{1/2}\del_{0}|_{\Omega^{1}(\S)\ot\Omega^{0}(S^{1})}}
     = (2\pi)^{-\c(\S)\dim\lt/2}\;\;.
\ee
This then reduces these determinants to those of the purely algebraic
operators (no derivatives) of the kind we encountered in the $G/G$ model.
The chiral nature of these determinants, which is not obvious from the
Chern-Simons point of view, arises because once one has chosen a metric on
$\S$ to implement the heat kernel regularization, one has in particular
chosen a complex structure on $\S$. It is then natural to decompose the
differential forms into their $(1,0)$ and $(0,1)$ parts. Note that this
is the way the (projective) dependence on the complex structure arises
quite generally in the canonical quantization of Chern-Simons theory.

\section{Evaluation of the Abelian theory and the Verlinde formula}

In this section we will evaluate the partition function
\be
Z_{\S}(S_{\phi F},k) = \int\! D\f\,DA\,
\det({\bf 1} - {\rm Ad}(e^{i\f}))^{\c(\S)/2}
\exp\left(\trac{i(k+h)}{2\pi}
\int_{\S}\tr\f F_{A}\right)\;\;,\label{zphif}
\ee
which, as we have seen in the preceding sections, is equal both to
the partition function of \CS\ theory on $\SS$ and that of the
$G/G$ model on $\S$.

\subsection{A Trivializing Map}

\noindent
This task can be simplified significantly by making use of a suitable
change of variables from $A$ to $F_{A}$, introduced in \cite{btbf1},
which trivializes the path integral and is in some sense an analogue
of the Nicolai map of supersymmetric field theories. In order to
implement this we will have to choose some gauge fixing condition
$G(A)=0$ for the Abelian gauge symmetry of the path integral
(\ref{zphif}). Here $G(A)$ is some scalar $\lt$-valued function
like e.g.~$G(A)=\del.A$ or $G(A)=n.A$. This will give rise to
a Faddeev-Popov determinant
\be
\Det \Delta_{FP} = \Det\left( \frac{\d\,G(A)}{\d A} d_{A}\right)\;\;.
\label{fp}
\ee
The next step is to perform the change of variables
\be
A \ra (F_{A},G(A))\;\;.
\ee
This maps the one-form $A$ to a pair of scalars and is well defined
because we have already eliminated the harmonic modes of $A$ (cf.~the
discussions on the Hodge decomposition in section 2 and the elimination
of the winding modes in section 5.5). It is easy to see that, for any
choice of $G(A)$, the Jacobian of this change of variables cancels
precisely against the Faddeev-Popov determinant (\ref{fp}),
\be
\Det \left(\frac{\d\,(F_{A},G(A))}{\d A}\right) = \Det\Delta_{FP}\;\;.
\ee
This is a manifestation of the fact that the Ray-Singer torsion is
trivial in even (and hence in particular in two) dimensions.
After performing the (trivial) integrals over $G(A)$ and its
multiplier field one is
then left with a path integral from which all derivatives have
disappeared,
\be
Z_{\S}(S_{\phi F},k) = \int\! D\f\,DF_{A}\,
\det(\dw(e^{i\f}))^{\c(\S)/2}
\exp\left(\trac{i(k+h)}{2\pi}
\int_{\S}\tr\f F_{A}\right)\;\;,\label{zphif2}
\ee
and which can be straightforwardly evaluated.

\subsection{Non-trivial $\bT$ Bundles and Integrality Conditions}

The only point that needs
some care is that the integral over $F_{A}$ should not extend over all
two-forms but only over those that arise as the curvature $F_{A}$ of
some connection $A$. As two-forms in two dimensions are automatically
closed, all that we need to require is the integrality condition
\be
\int_{\S}F^{l}_{A}\in 2\pi{\bf Z}\;\;\;\;\forall l = 1,\ldots,r=\dim\lt\;\;.
\ee
We do this by inserting the periodic delta function
\be
\d^{P}(\int_{\S}F^{l}_{A}) =
\sum_{n=-\infty}^{\infty}\exp\left(in\int_{\S}F^{l}_{A}\right)\label{sum}
\ee
into the path integral for each $l$ and can henceforth drop the (now
irrelevant)
label $A$ on $F_{A}$. Recalling that the label $l$ on $F^{l}$ refers
to an expansion of $A$ in terms of simple roots $\a_{l}$, and that the
fundamental weights $\la^{k}$ are dual to these (\ref{para}), we see
that we can write the sum over $r$-tuples
of integers arising from
(\ref{sum}) as a sum over elements $\la = \sum_{k}n_{k}\la^{k}$
of the weight lattice
\be
\Lambda = {\bf Z}[\la^{1},\ldots,\la^{r}]
\ee
of $\bG = SU(n)$. Thus (\ref{zphif2}) becomes
\be
(\ref{zphif2}) = \sum_{\lambda\in\Lambda}\int\! D\f\,DF\,
\det(\dw(e^{i\f}))^{\c(\S)/2}
\exp\left(\trac{i(k+h)}{2\pi}
\int_{\S}\tr\f F-i\int_{\S}\tr\lambda F\right)\;\;.\label{zphif3}
\ee
Let us briefly pause to explain the necessity of including non-trivial
$\bT$ bundles in this sum although the $\bG$ bundle we started off
with was (necessarily for simply connected $\bG$) trivial.
The reason for this is that any $\bT$ connection on $\S$, be it a
connection on a trivial bundle or not, can also
be regarded as a particular $\bG$ connection (technically speaking,
the structure group can always be extended from $\bT$ to $\bG$). Thus
an integral over connections on a trivial $\bG$ bundle will necessarily
have to include contributions from all the non-trivial $\bT$ sectors.
We encountered a similar phenomenon in section 5.4, where we saw that
an integral over maps from $\S$ to $\bG$, which are all homotopic to
the identity, devolved to an integral over maps to $\bT$ including
all non-trivial winding sectors (that these turned out not to contribute
to the path integral is tangential to the present discussion). Regarded
as maps from $\Sigma$ to $\bG$ which just happen to take their values
in $\bT\ss\bG$, these maps of course become contractible.

After this excursion we return to the path integral (\ref{zphif3}).
The integration over $F$ can now be performed, giving rise to a
delta function constraint on $\f$, so this step in our evaluation
of (\ref{zphif}) leads to
\bea
Z_{\S}(S_{\phi F},k) &=&
\sum_{\lambda\in\Lambda}\int\! D\f\,
\det(\dw(e^{i\f}))^{\c(\S)/2}
\delta(\trac{k+h}{2\pi}\f - \lambda) \nonumber\\
&=&
\prod_{k=1}^{r}\sum_{n_{k}}\int\! D\f\,
\det(\dw(e^{i\f}))^{\c(\S)/2}
\delta(\trac{k+h}{2\pi}\f_{k} - n_{k})
\;\;.\label{zphif4}
\eea
The first thing to note is that this equation implies in particular
that only the constant modes of $\f$ contribute to the partition
function. This has the two consequences mentioned above, namely a) that
the dilaton term of section 6.2 turns into the metric independent
Ray-Singer torsion of $\SS$, and b) that had we carried around the
kinetic term (\ref{g3}) for the compact scalars until now (instead of
eliminating it by the shift (\ref{g6})), it would disappear now.

One can gradually see the structure of the Verlinde
formula emerging. The correct integrand (which becomes a summand via
the delta function) has been around for some time. We have also
been forced to include a sum over the weight lattice, whose
summation range will be restricted by the compactness of $\f$.
In order to recognize the result as a sum over the highest
weights of integrable representations at level $k$, it is
convenient to restrict the integration range for $\f$ to a
fundamental domain of the action of the Weyl group $W$ on $\bT$
(which is the only piece of gauge freedom we have not yet fixed).
As $W$ is a finite group and the integral (\ref{zphif4}) is manifestly
$W$-invariant, the result will be the same as dividing the
integral by $|W|$, but not necessarily manifestly so, as the sum
will then extend over all the weights in the $W$-orbits of
highest weights of integrable representations. We postpone a discussion
of the overall normalization of the path integral, which is subject
to a standard renormalization ambiguity anyway, until after we have
convinced ourselves that up to normalization the partition function
is equal to the Verlinde formula.

\subsection{The Verlinde Formula for $SU(2)$}

\noindent
In this case, $\f$ is a single compact scalar, $r=1$ and
$\Lambda\sim{\bf Z}$ in (\ref{zphif4})
and the Weyl determinant (torsion) is $4\sin^{2}(\f/2)$. By the
action of the Weyl group the range of $\f$ is cut down from $[0,4\pi)$
to $[0,2\pi]$ and it is convenient to use the form of the Weyl integral
formula given in the last line of (\ref{w12}). Thus, for $SU(2)$ we
have
\be
Z_{\S}(S_{\f F},k) = \sum_{n=-\infty}^{+\infty}\int_{0}^{2\pi}\!d\f\,
\sin^{2-2g}(\f/2)\,\delta(\trac{k+2}{2\pi}\f - n)\;\;. \label{zphif5}
\ee
In particular, only certain discrete values of $\f$ contribute to the
path integral and due to the compactness of $\f$ only a finite number
of $n$'s give a non-vanishing contribution. Ignoring the boundary values
$n=0$ and $n=k+2$ for a moment (we will come back to them below) we
see that the allowed values of $\f$ are
\be
\f = \frac{2n\pi}{k+2}\;\;,\;\;\;\;\;\; n = 1,\ldots, k+1\;\;.
\ee
These points are in one-to-one correspondence with the $k+1$ integrable
representations of the $SU(2)$ WZW model at level $k$ and indeed we see
that, up to normalization, the partition function
\be
Z_{\S}(S_{\f F},k) = \sum_{n=1}^{k+1}\sin^{2-2g}(\trac{n\pi}{k+2})
\label{zphif6}
\ee
is precisely equal to the Verlinde formula
\be
\dim V_{g,k} = (\trac{k+2}{2})^{g-1}\sum_{j=0}^{k}
\left(\sin^{2}\trac{(j+1)\pi}{k+2}\right)^{1-g}\;\;.
\ee
It remains to come to terms with the values $\f=0$ and $\f=2\pi$,
arising from $n=0$ and $n=k+2$ and arising as
the boundary points of the reduced $\f$-range $[0,2\pi]$. The first
thing to note is, that these values correspond to the connections
on the circle with holonomy group $\{1\}$ (the trivial connection)
and $\{1,-1\}$ respectively. As such they are the most reducible
connections on the circle and require a special treatment in the
path integral. This can also be seen from our use of the Weyl integral
formula which, strictly speaking, only covers the regular elements
of $\bG$ or $\bT$, i.e.~excludes precisely the two special values
of $\f$ (for which the Weyl determinant vanishes).

The usual procedure would be
to either declare their contributions to be zero because of ghost zero modes
or to ignore these singular points.
Technically, this can be achieved by choosing the integration range for
$\f$ to be $[\eps,2\pi - \eps]$ and taking the limit $\eps\ra 0$.
This also takes care of the problem that these boundary values give rise
to infinities in the partition function in genus $g>1$ (as may be
seen from (\ref{zphif6})) and any other method of regulating these
infinities would also amount to ignoring these contributions.
We can take the attitude that the WZW models are defined by integrating
over fields with values in $\bG_{r}$. Such configurations are dense in
the space of fields and the path integral is naturally regularized
by the restriction to $\bG_{r}$.

As this procedure may nevertheless seem somewhat ad hoc, we want to
point out that there is also another reason for `dropping' the boundary values
and (more generally) the points on the boundary of the Weyl alcove. Namely,
as is well known there is a quantization ambiguity in Chern-Simons
theory (see e.g. \cite{axcs,imcs}), corresponding to the option to work
with $W$-even or $W$-odd wave functions. While both of these appear to lead
to perfectly unitary quantizations of \CS\ theory, it is only the latter
which turns out to be related to the current blocks of $G_{k}$ WZW models.
In particular, for $SU(2)$ this forces the wave functions to vanish at
$\f =0$ and $\f = 2\pi$. And, while the differences between the two
alternatives are quite far-reaching and subtle in general, for the
purposes of calculating the partition function they indeed only amount to
including or dropping the boundary values.

\subsection{The Verlinde Formula for $SU(n)$}

\noindent
The only complication that arises for $SU(n)$, $n>2$, is that we have
to prescribe a fundamental domain for the action of the Weyl group on
the torus $\bT = U(1)^{n-1}$. Alternatively, we are looking for a
fundamental domain of the action on $\lt$ of the semi-direct product of the
integral lattice (acting via translations) with the Weyl group (acting
via reflections). The advantage of this reformulation is that one now
recognizes this as a fundamental domain for the affine Weyl group (for
simply connected groups the integral lattice and the coroot lattice
coincide), which is known as a Weyl alcove or Stiefel chamber. In
particular, given such an alcove ${\bf P}$, we obtain a refinement of the
covering conjugation map (\ref{w1}) to a universal covering
\cite[Prop.~7.11]{btd}
\bea
\bG/\bT \times {\bf P} &\ra& \bG_{r}\;\;,\nonumber\\
(g,X)                  &\ra& \g \exp(X)\, g\;\;.
\eea
Hence this is an isomorphism if $\bG$ is simply connected and therefore
$\bf P$ is precisely the integration domain we require in the Weyl integral
formula instead of $\bT$ if we want to mod out by the Weyl group explicitly.

For $SU(n)$ such a Weyl alcove is determined by
$\a_{l}>0$ (fixing a Weyl chamber) and the one additional condition
$\sum\a_{l} < 2\pi$. As the fundamental weights are dual to the simple
roots, this amounts to the following conditions on
the integration range of $\f$:
\be
{\bf P} = \{\f_{l}:\,\f_{l}>0\,,\,\sum_{l=1}^{r=n-1}\f_{l}<2\pi\}\;\;.
\ee
Introducing this constraint into the path integral (\ref{zphif4}), one
finds that only those weights ($r$-tuples of integers) contribute
to the partition function which satisfy $n_{l}>0$ and $\sum n_{l} < k+n$,
i.e.~the allowed values of $\f$ are
\be
\f_{l} = \frac{2\pi n_{l}}{k+n}\;\;,\;\;\;\;\;\;n_{l}>0\;\;,\;\;
\sum n_{l} < k+n\;\;.
\ee
Again these are in one-to-one correspondence with the integrable
representations of the $SU(n)$ WZW model at level $k$ and, up
to an overall normalization, the partition function agrees with
the Verlinde formula given in the Introduction (with $\f = \lambda + \rho$).

\subsection{The Normalization of the Path Integral}

In the course of the derivation of the partition function
we have so far paid little attention to the various factors contributing to the
overall normalization.
And while it is certainly possible in principle to keep track of these,
it is also quite cumbersome. One might hence be tempted to leave it at that,
in particular when one keeps in mind that the normalization is subject to a
standard renormalization ambiguity $Z\ra v^{2-2g} Z$ arising from the
possibility to add terms $\sim\int R$ to the action without violating any
of the symmetries of the theory. However, the very fact that we are
calculating a dimension, which should at the very least be an integer,
forces us (and permits us) to go further than that. We will now sketch
how one could proceed.

The relation between the (naive)
partition function $Z_{g,k}$ of Chern-Simons theory at level $k$,
(\ref{zphif6}), and the dimension $\dim V_{g,k}$, consistent with
standard renormalization, is
\be
\dim V_{g,k} = a b^{g-1} Z(g,k)\;\;, \label{n1}
\ee
where we allow both $a$ and $b$ to depend on $k$. First of all,
it follows from the
fact that the moduli space of flat connections on the two-sphere is
a single point, that canonical quantization of Chern-Simons theory
on $S^{2}\times {\bf R}$ will give rise to a one-dimensional Hilbert
space. Demanding $\dim V_{0,k}=1$ and using $Z_{0,k} = (k+2)/2$
(as follows easily from (\ref{zphif6})), one finds that $b$ is
determined in terms of $a$ and (\ref{n1}) reduces to
\be
\dim V_{g,k} = a^{g}(\trac{k+2}{2})^{g-1}Z_{g,k}\label{n2}\;\;.
\ee
If we permitted ourselves to use the fact that in genus one the
dimension of the Hilbert space is equal to the number of
integrable representations, this would fix $a=1$ and lead to the
correct result. However, this fact relies not only on the knowledge that
the Hilbert space of Chern-Simons theory is the space of conformal
blocks of the WZW model, but also on the knowledge of that space's dimension.
And, in the spirit of this paper, we will
continue without making recourse to the connection between
\CS\ theory and conformal field theory.

By demanding that $\dim V_{1,k}$ be an integer one learns that
$a(k+1)$ has to be an integer. And one might suspect
that (\ref{n2}) cannot possibly be an integer for all $k$ and $g$ unless
$a$ itself is an integer. This is in fact correct and the remainder of
this section serves to establish just this (and to fix $a=1$).

First of all, one can e.g.~demand that
in the large $k$ limit $\dim V_{1,k}$ approach the volume
of the moduli space of flat connections. In genus one this moduli
space is two-dimensional and hence the determination of its
symplectic volume is elementary. As the symplectic form (or first
Chern class of the prequantum line bundle) is assumed to be $k$ times the
generator of the second cohomology, the volume (divided by $k$) is
just ${\rm Vol}(\M_{g=1}) =1$. This fixes $a(k+1)$ to be linear in $k$,
\be
\lim_{k\ra\infty}k^{-1}\dim V_{1,k}=1 \Rightarrow a = \frac{k+c}{k+1}\;\;,
\;\;\;\;c\in{\bf Z}
\ee
and leads to the somewhat improved expression
\be
\dim V_{g,k} = (\trac{k+c}{k+1})^{g}(\trac{k+2}{2})^{g-1}Z_{g,k}\;\;.
\label{n3}
\ee
This is as much as $g=0,1$ tell us. Furthermore, $k=1$ does not provide
any additional information in any genus, as the dimension comes out to
be $(c+1)^{g}$, which is an integer. However, by considering e.g.~genus 2
and successively calculating $\dim V_{2,k}$ from (\ref{n3}) for $k=2,4,\ldots$,
one finds that $c$ has to be of the form $c=1+3m$, that $m$ has to be of the
form $m=5n$, $\ldots$, leaving finally as the only possibility $c=1$.
This leads to the correctly normalized Verlinde formula.

While this line of argument was not very elegant (and other ways of fixing
the normalization are certainly also possible), its purpose was to illustrate
that, in principle, the correct normalization can be determined by elementary
methods. It remains a challenge to determine the regularization which
automatically leads to the correct normalization.

\subsection{Punctured Surfaces and the Fusion Rules}

On the basis of what we have achieved so far it turns out to be surprisingly
easy to derive the fusion rules, and more generally the dimension
$\dim V_{g,s,k}$ of the space of conformal blocks for a punctured Riemann
surface, from Chern-Simons theory or the $G/G$ model by explicit
calculation. These formulae arise as the expressions for correlation functions
of vertical Wilson loops (in Chern-Simons theory) or traces of $g$ (the
corresponding observables of the $G/G$ model)
directly in terms of the modular matrix $S$, as in \cite{ver,ms}.
We will again be content with illustrating this in the case of $SU(2)$.

Let $\c_{l}(h)$ be the character (trace) of $h\in SU(2)$
in the $(l+1)$-dimensional representation of $SU(2)$. We will only
consider integrable representations, $l\leq k$.
What we wish to prove is that the correlator
of $s$ such operators is given by the Verlinde formula for $\dim V_{g,s,k}$,
\be
\dim V_{g,s,k}(l_{1},\ldots,l_{s}) =
\langle\c_{l_{1}}(h)\ldots\c_{l_{s}}(h)\rangle_{g}\;\;.\label{f1}
\ee
As the characters are conjugation invariant, we need to know them only on the
maximal torus, where they can be expressed as
\be
\c_{l}(\f)\equiv\c_{l}(e^{i\f}) = \frac{\sin(l+1)\f/2}{\sin\f/2}\;\;.
\label{f2}
\ee
Repeating the steps of
sections 5 and 7.1-7.2 for this correlator instead of the partition function
one arrives at (\ref{zphif4}) with an insertion of $s$ characters in the
form (\ref{f2}). Then, just as in the case without insertions, evaluation
of the delta function will lead to a sum over the discrete
allowed values of $\f$ and one finds
\be
\langle\prod_{i=1}^{s}\c_{l_{i}}(\f)\rangle_{g} = (\trac{k+2}{2})^{g-1}
\sum_{j=0}^{k}\left(\sin\trac{(j+1)\pi}{k+2}\right)^{2-2g-s}\prod_{i=1}^{s}
\sin\trac{(j+1)(l_{i}+1)\pi}{k+2}\;\;.\label{f3}
\ee
It is now convenient to introduce the $(k+1)\times(k+1)$ matrix $S$,
\be
S_{ij} = (\trac{2}{k+2})^{1/2}\sin\trac{(i+1)(j+1)\pi}{k+2}\;\;.\label{f4}
\ee
This symmetric and orthogonal
matrix is usually introduced as the modular matrix which implements the
modular transformation $\tau\ra -1/\tau$ on the genus 1 conformal blocks
(Weyl-Kac characters),
\be
\c_{i}^{wk}(-1/\tau) = \sum_{j} S_{ij}\c_{j}^{wk}(\tau)\;\;.
\ee
We see that within our formalism it arises quite naturally as well.
In terms of $S$ the above correlator is
\be
\langle\prod_{i=1}^{s}\c_{l_{i}}(\f)\rangle_{g} =
\sum_{j=0}^{k} (S_{j0})^{2-2g-s}\prod_{i=1}^{s}S_{jl_{i}}\;\;.
\ee
This is precisely the formula obtained in \cite{ver,ms}
for the dimension of
the space of conformal blocks on a surface of genus $g$ with punctures
labelled by the representations $\{l_{i}\}$. In particular, for the
two- and three-point functions on the sphere we obtain
\bea
\langle\c_{l}(\f)\c_{m}(\f)\rangle_{g=0} &=& \d_{lm}\nonumber\\
\langle\c_{l}(\f)\c_{m}(\f)\c_{n}(\f)\rangle_{g=0} &=&
\sum_{j=0}^{k}\frac{S_{jl}S_{jm}S_{jn}}{S_{j0}}\equiv N_{lmn}\;\;.
\eea
It is interesting to note that in this derivation the fusion rules
appear naturally in already diagonalized form. In particular, once
the delta function constraint on $\f$ has been imposed, the characters
appear only in the form
\be
\c_{l}^{(j)} = \frac{S_{lj}}{S_{0j}}\;\;.
\ee
These `discrete characters' are the eigenvalues of the fusion matrix
$(N_{l})_{mn}=N_{lmn}$ and are known to satisfy the fusion rules
all on their own,
\be
\c_{l}^{(j)}\c_{m}^{(j)} = N_{lmn}\c_{n}^{(j)}\;\;.
\ee

\subsection*{Acknowledgements}

We thank M. Crescimanno, R. Dijkgraaf, K.S. Narain, M.S. Narasimhan and
T.R. Ramadas for comments, criticism and discussions at various stages in this
work, and S. Randjbar-Daemi for urging us to make things as simple as
possible.

\rnc{\Large}{\normalsize}

\end{document}